\documentclass[review,11pt]{elsarticle}
\usepackage{bm,amsmath,amssymb,epsfig,dsfont}
\usepackage{amsthm}
\usepackage{algpseudocode}
\usepackage{url}
\usepackage{algorithm}
\usepackage[nofiglist, nomarkers]{endfloat}
\usepackage{color,soul}
\usepackage{graphicx}

\newtheorem{Lem1}{Lemma}

\newtheorem{Thr1}{Theorem}

\journal{Signal Processing}

\begin{document}

\begin{frontmatter}


\title{Optimal Transmission Policy for Cooperative Transmission with Energy Harvesting and Battery Operated Sensor Nodes}

\author{Lazar~Berbakov\corref{cor1}}
\ead{lazar.berbakov@cttc.es}
\author{Carles~Ant\'{o}n-Haro}
\ead{carles.anton@cttc.es}
\author{Javier~Matamoros}
\ead{javier.matamoros@cttc.es}
\cortext[cor1]{Corresponding author, Phone +34 936452918 Ext: 2216}
\address{Centre Tecnol\`{o}gic de Telecomunicacions de Catalunya\\
            Parc Mediterrani de la Tecnologia (PMT) - Building B4\\
            Av. Carl Friedrich Gauss 7\\
            08860 - Castelldefels (Barcelona) - Spain}


\begin{abstract}
In this paper, we consider a scenario where one energy harvesting
and one battery operated sensor cooperatively transmit a common
message to a distant base station. The goal is to find the
\emph{jointly} optimal transmission (power allocation) policy which
maximizes the total throughput for a given deadline. First, we
address the case in which the storage capacity of the energy
harvesting sensor is \emph{infinite}. In this context, we identify
the necessary conditions for such optimal transmission policy. On
their basis, we first show that the problem is convex. Then we go
one step beyond and prove that (i) the optimal power allocation for
the energy harvesting sensor can be computed independently; and (ii)
it unequivocally determines (and allows to compute) that of the
battery operated one. Finally, we generalize the analysis for the
case of \emph{finite} storage capacity. Performance is assessed by
means of computer simulations. Particular attention is paid to the
impact of finite storage capacity and long-term battery degradation
on the achievable throughput.
\end{abstract}

\begin{keyword}
wireless sensor networks \sep energy harvesting \sep cooperative
transmission


\end{keyword}

\end{frontmatter}


\vspace{-0.3cm}
\section{Introduction}
\vspace{-0.1cm} Sensor nodes are typically powered by batteries
that, quite often, are either costly, difficult or simply impossible
to replace. Clearly, this limits network lifetime. Energy harvesting
makes it possible to overcome this drawback by allowing sensors to
harvest energy from e.g. solar, mechanical, or thermal sources. The
harvested energy is typically stored in a device (e.g. battery,
super capacitor) and then supplied for communication and/or
processing tasks when needed. In recent years, many authors have
analyzed how to optimally use such harvested energy. For
\emph{single}-sensor scenarios, in \cite{Yang_Ulukus_2011} the
authors derive the optimal transmission policy which minimizes the
time needed to deliver all data packets to the destination subject
to causality constraints on energy and packet arrivals. In
\cite{Tutuncuoglu_Yener_2012}, the authors go one step beyond and,
unlike \cite{Yang_Ulukus_2011}, they consider \emph{finite} storage
capacity effects. In both cases, the energy harvesting instants and
amounts of energy harvested are assumed to be known a priori. Ozel
\emph{et al} generalize the analysis to Rayleigh-fading channels and
for the case in which the information on the harvested energy and
channel gains is either causally or non-causally known
\cite{Ozel_Tutuncuoglu_2011_Journal}. Other works in the literature
have addressed scenarios with \emph{multiple} energy harvesting
terminals. This includes studies for the multiple-access
\cite{Yang2012_JCN}, interference \cite{Tutuncuoglu_Yener_2011_JCN},
relay \cite{Gunduz_Devillers_2011} and {broadcast}
\cite{Yang2012_ToWC, Ozel2012_ToWC} channels.

Distributed beamforming techniques allow nodes in a Wireless Sensor
Network (WSN) to act as a virtual antenna array in order to reach a
distant Base Station (BS) or data sink. This, however, requires
accurate frequency and phase synchronization over sensors. To that
aim, one can resort to the iterative synchronization scheme with
one-bit of feedback proposed in \cite{Mudumbai2010}, or
opportunistic sensor selection schemes \cite{Pun2009}.

In this paper, we consider a scenario where one energy harvesting
(EH) and one battery operated (BO) sensor cooperate to transmit
(beamform) a common message to a distant base station. This differs
from the scenarios in \cite{Yang2012_JCN,
Tutuncuoglu_Yener_2011_JCN, Gunduz_Devillers_2011}, where
\emph{both} sensors/terminals had energy harvesting capabilities.
Our goal is to find the \emph{jointly} optimal power allocation
strategy which maximizes the total throughput for a given deadline
(e.g. the time by which batteries could be replaced). This problem
is equivalent to the one addressed e.g. in \cite{Yang_Ulukus_2011,
Ozel_Tutuncuoglu_2011_Journal} but here we consider the more general
case with \emph{multiple} transmitters. Besides, and unlike the
Multiple-Access Channel (MAC) scenarios in \cite{Yang2012_JCN},
sensors here attempt to convey a \emph{common} message to the
destination. We also go one step beyond the distributed beamforming
approaches in \cite{Mudumbai2010, Pun2009} where, implicitly, all
sensors were assumed to be battery operated, and investigate the
impact of \emph{energy harvesting} constraints on performance. As in
\cite{Yang2012_JCN}, we initially assume that the energy harvesting
sensor is equipped with a re-chargeable battery of \emph{infinite}
storage capacity. In this context, we identify the necessary
conditions for the jointly optimal transmission policy. This leads
to a problem that we show to be convex. Furthermore, and as an
extension to our previous work in \cite{Berbakov2012_ISWCS}, we
prove that the optimal policy for the EH node can be computed
independently from that of the BO one, and propose an algorithm to
compute the latter from the former. Next, we generalize the analysis
for a scenario in which, as in \cite{Tutuncuoglu_Yener_2012}, the
storage capacity of the EH sensor is \emph{finite}. We also consider
imperfections in the re-chargeable battery of the EH sensor. More
specifically, we focus on the impact of \emph{long-term} capacity
degradation, as opposed to the (short-time) battery leakage effects
addressed in \cite{Devillers2012_JCN}
\vspace{-0.3cm}
\section{Signal and Communication Model}
\vspace{-0.1cm} \label{sec:signal_model}
%
Two sensors cooperate to transmit a common message $m(t)$ to a
distant base station. The received signal thus reads
\vspace{-0.1cm}
\begin{equation}\label{eq:received_signal}
    r(t)=m(t)\left(\sum_{i=1}^{2} w_i(t) e^{j\psi_i(t)}\right) + n(t)
\end{equation}
where the common message is given by {$m(t)=\sum_l x_l g(t-l T_s)$,
with $\{x_l\}$ standing for a sequence of zero-mean complex Gaussian
symbols with unit variance ($T_s$ is the symbol period) and $g(t)$
denoting the impulse response of a bandlimited pulse (unit
bandwidth)}; $w_i(t)=\sqrt{p_i(t)}e^{j\phi_i(t)}$ denotes the
time-varying complex transmit weights in polar notation (to be
designed); $e^{j\psi_i(t)}$ stands for the phase shift of the
(Gaussian) sensor-to-base station channels; and $n(t)$ is zero-mean
complex additive white Gaussian noise with unit variance (i.e.
$n(t)\sim\mathcal{CN}(0,1))$. In the sequel, we assume that by
properly designing $\phi_i(t)$ the channel phase and, where
relevant, oscillator offsets can be ideally pre-compensated
\cite{Mudumbai2010}. Frequency and time synchronization is assumed,
as well. Hence, the sensor network behaves as a virtual antenna
array capable of \emph{beamforming} the message to the base station.
Without loss of generality, we let the first sensor be the one with
energy harvesting capabilities, and the second to be battery
operated. Consequently, we hereinafter denote by $p^H\!(t)\triangleq
p_1(t)$ and $p^B\!(t)\triangleq p_2(t)$ the transmit power at the
energy \emph{harvesting} and \emph{battery} operated sensors,
respectively. Bearing all this in mind, the instantaneous received
power at the base station is given by $p_{BF}\!(t) =
(\sqrt{p^H\!(t)} + \sqrt{p^B\!(t)})^2$. The total throughput for a
given deadline $T$ then reads
\vspace{-0.1cm}
\begin{equation}
\label{eq.throughput_definition_general} G_T(p^H\!(t),p^B\!(t)) =
\int_0^T \log \left(1+p_{BF}(t)\right) dt.
\end{equation}
Our goal is to find the \emph{jointly} optimal transmission (power
allocation) policies $p^H\!(t)$ and $p^B\!(t)$ such that $G_T$ is
maximized subject to the causality constraints imposed by the energy
harvesting process, namely\footnote{For scenarios where the storage
capacity of the EH sensor is finite, additional constraints must be
introduced (see Section \ref{sec.Finite_capacity}).},
\vspace{-0.1cm}
\begin{eqnarray}
e^H\!(t) \leq E^H\!(t) &\triangleq&  \sum_{k:s_k < t} E_k^1\label{eq.power_allocation_general_constraints_1}\\
e^B\!(t)  \leq E^B\!(t) &\triangleq&  E_0^2,
\label{eq.power_allocation_general_constraints_2}
\end{eqnarray}
\begin{figure}[t]
   \centering
   \includegraphics[width=0.99\columnwidth]{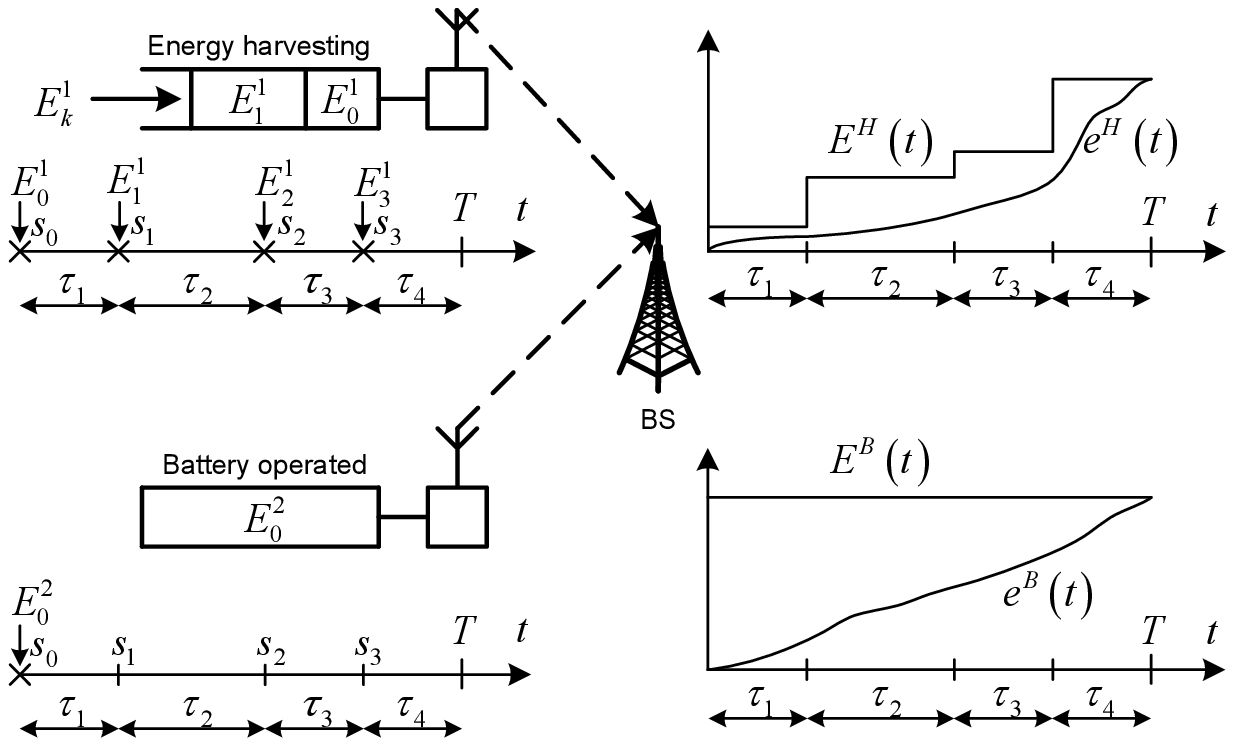}
      \vspace{-0.2cm}
   \caption{Network with energy harvesting and battery operated sensors (left); cumulative energy constraints and energy consumption curves (right).}
   \vspace{-0.3cm}
   \label{fig.EH_WSN_system_setup}
\end{figure}
for $0 \leq t \leq T$, where $e^H\!(t)=\int_0^t p^H\!(\tau) d\tau $
and $e^B\!(t)=\int_0^t p^B\!(\tau) d\tau$ denote the energy
\emph{consumption} (EC) curves; and $E^H\!(t)$, $E^B\!(t)$ stand for
the cumulative energy \emph{harvesting} (cEH) constraints (see Fig.
\ref{fig.EH_WSN_system_setup}). In the above expression, $E_k^i$
accounts for the amount of energy harvested by sensor $i$ in the
$k^\text{th}$ {event} ($k=0\ldots N-1$). We define \emph{event}
$s_k$ as the time instant in which some energy is harvested by
\emph{any} of the sensors in the network ($E_k^i=0$ for the sensor
not harvesting any energy in that event). Both the events and the
amounts of energy harvested $E_k^i$ are assumed to be known a
priori. Further, we impose $E_0^i>0$ for all $i$ (sensors) so that
collaborative transmission can start immediately, that is, from
$t=0$. For battery operated sensors, we have $E_k^i=0$ for $k>0$
and, thus, the cumulative energy harvesting function is constant for
the whole period. For the EH sensor, on the contrary, it is given by
a staircase function. Finally, we define \emph{epoch} as the time
elapsed between two consecutive events $s_k$ and $s_{k-1}$. Its
duration is given by $\tau_k \triangleq s_k - s_{k-1}$ for
$k=1\ldots N-1$ and, likewise, we define $\tau_{N} \triangleq T -
s_{N-1}$. A given transmission policy is said to be \emph{feasible}
(yet, perhaps, not optimal) if, as imposed by
\eqref{eq.power_allocation_general_constraints_1} and
\eqref{eq.power_allocation_general_constraints_2}, the energy
consumption curves lie below cumulative energy harvesting ones at
all times (or occasionally hit them).
\vspace{-0.3cm}
\section{Necessary conditions for the optimality of the transmission policy}
\vspace{-0.1cm}
%
The following lemmas give the necessary optimality conditions under
the assumption of \emph{infinite} storage capacity for the EH
sensor. Moreover, the insights gained into the problem structure
allow us to compute the jointly optimal transmission policies in
Section \ref{sec.OptimalPowerAllocationPolicy}. Unless otherwise
stated, the lemmas hold for \emph{both} the energy harvesting and
battery operated sensors.
\begin{Lem1}
\label{Lemma2} The transmit power in each sensor remains constant
between consecutive events.
\end{Lem1}
In other words, the power/rate in \emph{each} sensor only
potentially changes when new energy arrives to \emph{any} of them
\footnote{In our scenario, only one sensor harvests energy. Still,
this lemma holds for a more general case with multiple energy
harvesters.}. The proof of this lemma, which is based on Jensen's and
Cauchy's inequalities, can be found in Appendix A. This lemma
implies that \mbox{$p^H\!(t) = p^H_k, p^B\!(t) = p^B_k$ for $s_{k-1}
\leq t < s_{k}$}. That is, the power allocation curves $p^H\!(t)$
and $p^B\!(t)$ are necessarily staircase functions and, hence, the
energy consumption curves $e^H\!(t)$ and $e^B\!(t)$ are piecewise
linear. This observation allows us to pose the original problem
\eqref{eq.throughput_definition_general} in a convex optimization
framework in which a numerical (or analytical) solution is easier to
find. This will be accomplished in the next section.
\begin{Lem1}
\label{Lemma1} All the harvested/stored energy must be consumed by
the given deadline $T$.
\end{Lem1}
This means that, necessarily, the cumulative energy harvesting
curves reach the energy consumption constraints at time instant $T$.
\begin{proof}[Proof] Lemma \ref{Lemma1} can be easily proved by
contradiction. Assume that the optimal transmission policy does not
fulfill such condition. We could think of a \emph{feasible} policy
such that (i) the set of curves $e^H\!(t)$ and $e^B\!(t)$ differ
from the \emph{optimal} ones in the last epoch only, namely, for
$t\in [s_{N-1} \ldots T)$; and (ii) it verifies $e^H\!(T)=E^H\!(T)$
and $e^B\!(T)=E^B\!(T)$. Being {piecewise linear (and continuous)},
these curves would necessarily lie \emph{above} the optimal ones at
least in part of such last epoch, this resulting in a higher
received power and throughput. This contradicts the optimality of
the original transmission policy.
\end{proof}
\begin{Lem1}
\label{Lemma3} If feasible, a transmission policy with constant
transmit power in each sensor between any two (i.e. not necessarily
consecutive) events turns out to be optimal {for the period of time
elapsed between these two events}.
\end{Lem1}
This lemma goes one step beyond and states that Lemma \ref{Lemma2}
also holds for \emph{non-consecutive} events, as long as a constant
transmit power policy in \emph{both} sensors is feasible\footnote{In
our setting, this can only be constrained by the cEH curve of the EH
sensor.} for this period. This follows directly from the proof of
Lemma \ref{Lemma2} but, since one or more energy harvests might take
place in between the initial and final events, feasibility needs to
be ensured (clearly, this is not needed in Lemma \ref{Lemma2}).
\begin{Lem1}
\label{Lemma4} The transmit powers for an energy harvesting sensor
with infinite storage capacity are monotonically increasing, i.e.
$p^H_1 \leq p^H_2 \leq \ldots \leq p^H_N$
\end{Lem1}
\begin{figure}[t]
   \centering
   \includegraphics[width=0.99\columnwidth]{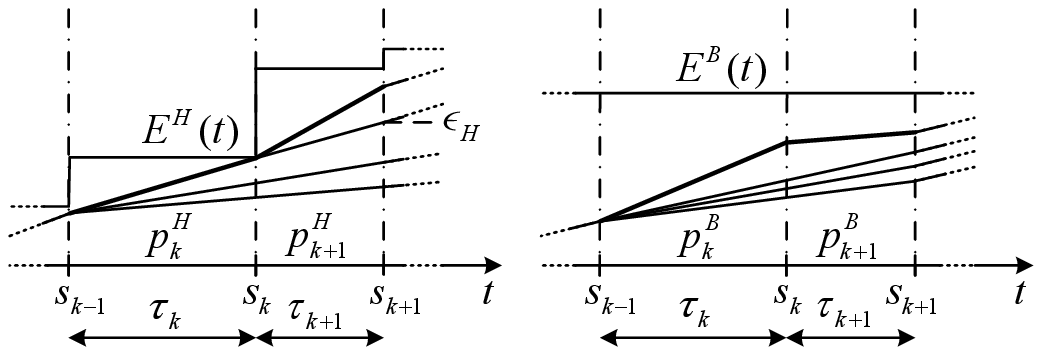}
      \vspace{-0.2cm}
   \caption{Transmit powers at the EH sensor are monotonically increasing (Lemma \ref{Lemma4}).}
   \vspace{-0.3cm}
   \label{fig.power_changes}
\end{figure}
\begin{proof}[Proof]
This property follows from the fact that $E^H\!(t)$ is a staircase
function. Assume that the power allocation policy before $s_{k-1}$
and after $s_{k+1}$ is optimal. As shown in Fig.
\ref{fig.power_changes}, the optimal EC curve verifies
$e^H\!(s_{k+1})\in \left(e^H\!(s_{k-1}),E^H\!(s_{k+1})\right]$. For
$e^H\!(s_{k+1})\in \left(e^H\!(s_{k-1}),\epsilon_H\right]$, we know
from Lemma \ref{Lemma3} that a constant power allocation for the
energy harvesting \emph{and} battery operated sensors turns out to
be optimal for $\left[s_{k-1}, s_{k+1}\right)$ (and, hence, for
$\left[0,T\right]$). In particular, this implies that $p^H_{k+1} =
p^H_{k}$. For $e^H\!(s_{k+1})\in \left(\epsilon_H,
E^H\!(s_{k+1})\right]$, on the contrary, the fact that $e^H\!(t)$ is
continuous and piecewise linear can only be ensured if (and only if)
$p^H_{k+1}> p^H_{k}$. By repeatedly applying this reasoning to all
consecutive epoch pairs the proof follows. As for the relationship
between $p^B_{k+1}$ and $p^B_{k}$, nothing can be said yet. Still,
the fact that $E^B\!(t)$ is a \emph{constant} function does not
impose any additional restrictions to the power allocation policy of
the BO sensor in $\left[s_{k-1}, s_{k+1}\right]$.
\end{proof}
\begin{Lem1}
\label{Lemma5} Transmit powers are strictly positive.
\end{Lem1}
\begin{figure}[t]
   \centering
   \includegraphics[width=0.99\columnwidth]{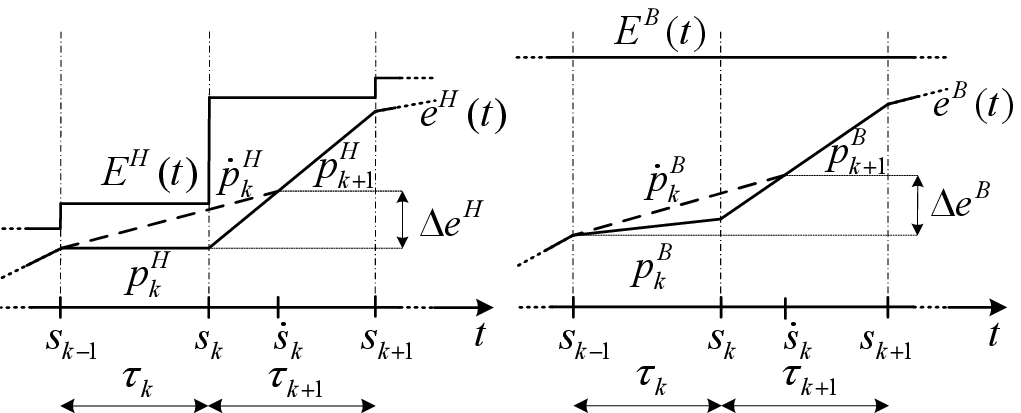}
      \vspace{-0.2cm}
   \caption{Transmit powers are strictly positive (Lemma \ref{Lemma5})}
   \vspace{-0.3cm}
   \label{fig.power_positive}
\end{figure}
\begin{proof}[Proof]
Again, this can be proved by contradiction. Assume that the power
allocation policy before $s_{k-1}$ and after $s_{k+1}$ is optimal.
Assume that, as shown in Fig. \ref{fig.power_positive}, the optimal
policy for the $\left[s_{k-1}, s_{k+1}\right)$ period verifies
$p^H_{k}=0$ and $p^H_{k+1}>0$. One could think of a new (and
feasible) transmission policy given by $\dot{p}^H_{k}=\frac{\Delta
e^H}{\dot{s}_k-s_{k-1}}$ and $\dot{p}^B_{k}=\frac{\Delta
e^B}{\dot{s}_k-s_{k-1}}$ for $t\in \left[s_{k-1},\dot{s}_k\right)$;
and $\dot{p}^H_{k+1}=p^H_{k+1}$ along with
$\dot{p}^B_{k+1}=p^B_{k+1}$ for $t\in
\left[\dot{s}_k,s_{k+1}\right)$. From the proof\footnote{Although
Lemma \ref{Lemma3} holds for EH \emph{events}, its \emph{proof} has
a broader scope and encompasses \emph{any} time instant, such as
$s_k'$. See Appendix A.} of Lemma \ref{Lemma3}, we know that the new
policy achieves higher throughput than the original one in
$\left[s_{k-1},\dot{s}_k\right)$ and, thus, in
$\left[s_{k-1},s_{k+1}\right)$ too. Yet not optimal (since this new
policy e.g. contradicts Lemma \ref{Lemma2}), this proves that the
original policy given by $p^H_{k}=0$ and $p^H_{k+1}>0$ was not
optimal either. Along the same lines, one can easily find a new
feasible policy achieving higher throughput than $p^B_{k}=0$ and
$p^B_{k+1}>0$ for the battery operated sensor.
\end{proof}
\vspace{-0.3cm}
\section{Computation of the optimal transmission policy}
\vspace{-0.1cm} \label{sec.OptimalPowerAllocationPolicy}
Lemma \ref{Lemma2} allows us to re-write the original optimization
problem given by the score function
\eqref{eq.throughput_definition_general} and the causality
constraints \eqref{eq.power_allocation_general_constraints_1} and
\eqref{eq.power_allocation_general_constraints_2} as follows (to
recall, our focus here is on scenarios where the storage capacity of
the EH sensor is \emph{infinite}):
\vspace{-0.1cm}
\begin{eqnarray}
&&\underset{\{p^H_k\}_{k=1}^{N},\{p^B_k\}_{k=1}^{N}}{\text{max}}  \sum_{k=1}^{N} \tau_k \log \left(1+(\sqrt{p^H_k} +\sqrt{p^B_k})^2\right)\label{eq:score_function}\\
&&\text{    s.t.:}\nonumber\\
&&\quad \sum_{k=1}^n \tau_k p^H_k \leq E_n^H =  \sum_{k=0}^{n-1} E_k^1 \; \text{  for}\; n=1\ldots N \label{eq_constraint_H}\\
&& \quad \sum_{k=1}^n \tau_k p^B_k \leq E_n^B =  E_0^2 \; \text{  for}\; n=1\ldots N \label{eq_constraint_B}\\
&&\quad p^H_k > 0 \; \text{  for}\; k=1\ldots N\\
&&\quad p^B_k > 0 \; \text{  for}\; k=1\ldots N
\label{eq.optimization_problem_1}
\end{eqnarray}
where we have defined $E_n^{H} \triangleq E^{H}(t)$ for
$t\in[s_{n-1}, s_{n})$, and where the last two strict inequalities
follow from Lemma \ref{Lemma5}. The problem is convex since all the
constraints are affine and linear, and the objective function is
concave, as we will prove next. To that aim, we observe that the
$k^{\text{th}}$ term in the summation exclusively depends on the
corresponding optimization variables $p^H_k$ and $p^B_k$ (i.e. no
cross-term variables). Hence, it suffices to show that an arbitrary
term in the summation, namely, $G_1(p^H,p^B) = \tau \log
(1+(\sqrt{p^H} + \sqrt{p^B})^2)$ is concave (indices have been
omitted for brevity). Or, alternatively, that
$G_2(p^H,p^B)=-G_1(p^H,p^B)$ is convex. The latter can be verified
by realizing that, for $p^H>0$ and $p^B>0$, its $2 \times 2$ Hessian
is positive definite, namely, \mbox{$\nabla^2 G_2(p^H,p^B) \succ
0$}. Since the optimization problem is strictly convex, its unique
solution can at least be found numerically (e.g. by resorting to
interior point methods). However, this task is computationally
intensive, in particular when the number of energy harvesting events
$N$ is large. This motivates the following lemma and two theorems
from which a semi-analytical and less computationally intensive
solution to the optimization problem can be obtained.
\begin{Lem1}
\label{Lemma6} The jointly optimal power allocation policy is such
that, whenever the transmit power changes, the energy consumed by
the energy harvesting sensor up to that time instant, equals the
energy harvested by such sensor up to that instant {(i.e, the stored
energy is zero)}.
\end{Lem1}
The proof of this Lemma is based on the Karush-Kuhn-Tucker (K.K.T.)
conditions associated to the (joint) optimization problem
\eqref{eq:score_function}-\eqref{eq.optimization_problem_1}. Details
can be found in Appendix B.

\vspace{0.3 cm}

The next theorems state the main result of this paper since they
allow to effectively compute the optimal transmissions policies for
the EH and BO sensors, respectively.
\begin{Thr1}
The optimal transmission policy for the energy harvesting sensor,
$\{\breve{p}^{H}_k\}_{k=1}^N$, can be computed independently from
that of the battery operated one. The associated energy consumption
curve turns out to be the shortest string {starting in $t=0$, ending
in $t=T$, and lying} below the cumulative energy harvesting curve.
\label{Th:policy_energy_harvester}
\end{Thr1}
\begin{figure}[t]
   \centering
   \includegraphics[width=0.99\columnwidth]{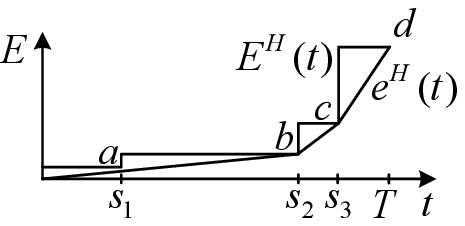}
      \vspace{-0.2cm}
   \caption{The optimal energy consumption curve for the EH sensor is given by the shortest
string below the cumulative EH curve (Theorem
\ref{Th:policy_energy_harvester})}
   \vspace{-0.3cm}
   \label{fig.Theorem_1}
\end{figure}

\begin{proof}[Proof]
As we will prove next, Lemmas \ref{Lemma2} to \ref{Lemma6}
unequivocally determine the optimal transmission policy for the EH
sensor. First note that, in order to satisfy the energy causality
constraint, the corresponding EC curve must lie below the cEH curve.
From Lemma \ref{Lemma1}, it follows that the EC curve reaches the
cEH curve at $t=T$. Besides, from Lemmas \ref{Lemma2} and
\ref{Lemma3}, we know that the transmit power only potentially
changes at the energy harvesting events. Consequently, the optimal
EC curve must be linear between them (i.e. piecewise linear).
Moreover, Lemma \ref{Lemma6} dictates that, whenever the transmit
power (slope) changes at an energy harvesting event, the EC curve
hits the cEH curve. Based on these facts, we conclude that the first
linear part of EC curve must connect the origin with some corner
point on the cEH curve (see Fig \ref{fig.Theorem_1}). Because of
Lemma \ref{Lemma4}, we must choose the one with the minimal slope,
since otherwise the constraint on energy causality (point $c$) or
monotonicity property of Lemma \ref{Lemma4} (point $a$) would not be
satisfied. Clearly, in Fig \ref{fig.Theorem_1} this corresponds to
point $b$. Once this point is identified, the algorithm can be
iteratively applied until we find the optimal policy until deadline
$T$. As a result, the EC curve is given by the shortest string below
the cEH curve. It must be noted that this algorithm is equivalent to
the one presented in \cite{Yang_Ulukus_2011}. However, the
interesting points are that (i) we have proved that it continues to
be optimal in a scenario where \emph{two} sensors, one of them
battery operated, collaborate to send the message (vs. one sensor in
\cite{Yang_Ulukus_2011}); and that (ii) no information on the BO
sensor (i.e. its optimal EC curve) is needed to determine it.
\end{proof}
\begin{Thr1}
\label{Th:policy_battery_operated} Upon finding the optimal
transmission policy for the energy harvesting sensor, the optimal
transmission policy for the battery operated one,
$\{\breve{p}^{B}_k\}_{k=1}^N$, can be computed with the iterative
procedure given by Algorithm \ref{alg_battery}.
\end{Thr1}
\begin{algorithm}[t]
\caption{Optimal policy for the battery operated
sensor}\label{alg_battery}
\begin{algorithmic}[1]
\State choose some small $\delta > 0$ \Comment{Step for searching}
\State $m := 0$ \Comment{Iteration index} \State $E^B_{T}:=E_0^2$
\Comment{Energy stored in the battery} \Repeat \State $m := m+1$
\ForAll {$k=1\ldots N$} \State $B_{k,m} := m \delta$, \State
\text{solve} $\breve{p}^H_k = \frac{B_{k,m}(A_{k,m} + B_{k,m} -
A_{k,m} B_{k,m})}{A_{k,m} (A_{k,m} + B_{k,m})^2}$ \text{for}
$A_{k,m}$ \State $p^B_{k,m} \gets
\left(\frac{A_{k,m}}{B_{k,m}}\right)^2 \breve{p}^H_k$ \EndFor \State
$E^B_{T,m} := \sum_k \tau_k p^B_{k,m}$ \Until{$E^B_{T,m} = E^B_{T}$}
\State $ \breve{p}^B_k \gets p^B_{k,m}$ $\forall k$
\end{algorithmic}
\end{algorithm}
\begin{proof}[Proof]
This algorithm stems from the proof of Lemma \ref{Lemma6} in
Appendix B (see \emph{Remark}). The real-valued variables $A_k$ and
$B_k$ (or their counterparts for iteration $m$, namely, $A_{k,m}$
and $B_{k,m}$) are linear functions of the Lagrange multipliers
associated to the constraints \eqref{eq_constraint_H} and
\eqref{eq_constraint_B}, respectively. Therefore, the equation in
Step 8 provides a connection between the primal and dual solutions
of the problem. Since $\breve{p}^H_k$ is already known from Theorem
\ref{Th:policy_energy_harvester}, for each value of $B_{k,m}$ to be
tested (from Appendix B we know that all the $B_k$s are identical
and equal to the largest Lagrange multiplier associated to
\eqref{eq_constraint_B}, which is enforced in Step 7), the
associated $A_{k,m}$ can be found by solving the corresponding third
order equation (a single real-valued root exists). From
$\breve{p}^H_k$, $A_{k,m}$, and $B_{k,m}$, an estimate of the
optimal transmission policy for the battery operated sensor for the
current iteration, namely, $\{p^{B}_{k,m}\}_{k=1}^N$, follows in
Step 9. If the total energy consumed until time instant $T$ by the
battery operated sensor, computed in Step 9, equals the energy
(initially) stored in it, $E^B_{T}$, the iterative algorithm stops.
The stopping condition not only ensures that Lemma \ref{Lemma1} is
fulfilled but also, it implies that the \emph{whole} transmission
policy for the battery operated sensor $\{\breve{p}^{B}_k\}_{k=1}^N$
is feasible. In summary, we have found the optimal transmission
policy for the BO sensor by (i) conducting a grid search over one
variable of the dual solution
; and (ii) checking in each iteration whether the
unknown part of the primal solution results from the algorithm is
feasible. Clearly, the choice of $\delta$ leads to a number of
trade-offs in terms of accuracy and number of iterations needed.
\end{proof}

{As for algorithmic convergence, one can easily prove that each
element in the set of transmit powers $\{{p}^{B}_k\}_{k=1}^N$ is a
\emph{monotonically decreasing} function in $\nu_N$ (the only
non-zero element in the dual solution, see Appendix B for details).
Likewise, $E_T^B$ is a monotonically decreasing function in $\nu_N$
as well. In other words, there exists a \emph{one-to-one mapping}
function between the primal and dual solutions. This turns out to be
a sufficient condition for the algorithm to converge, as long as a
\emph{sufficiently} small step size $\delta$ is used for the grid
search over some range of $\nu_N$ values.}

\begin{figure}[t]
   \centering
   \includegraphics[width=0.99\columnwidth]{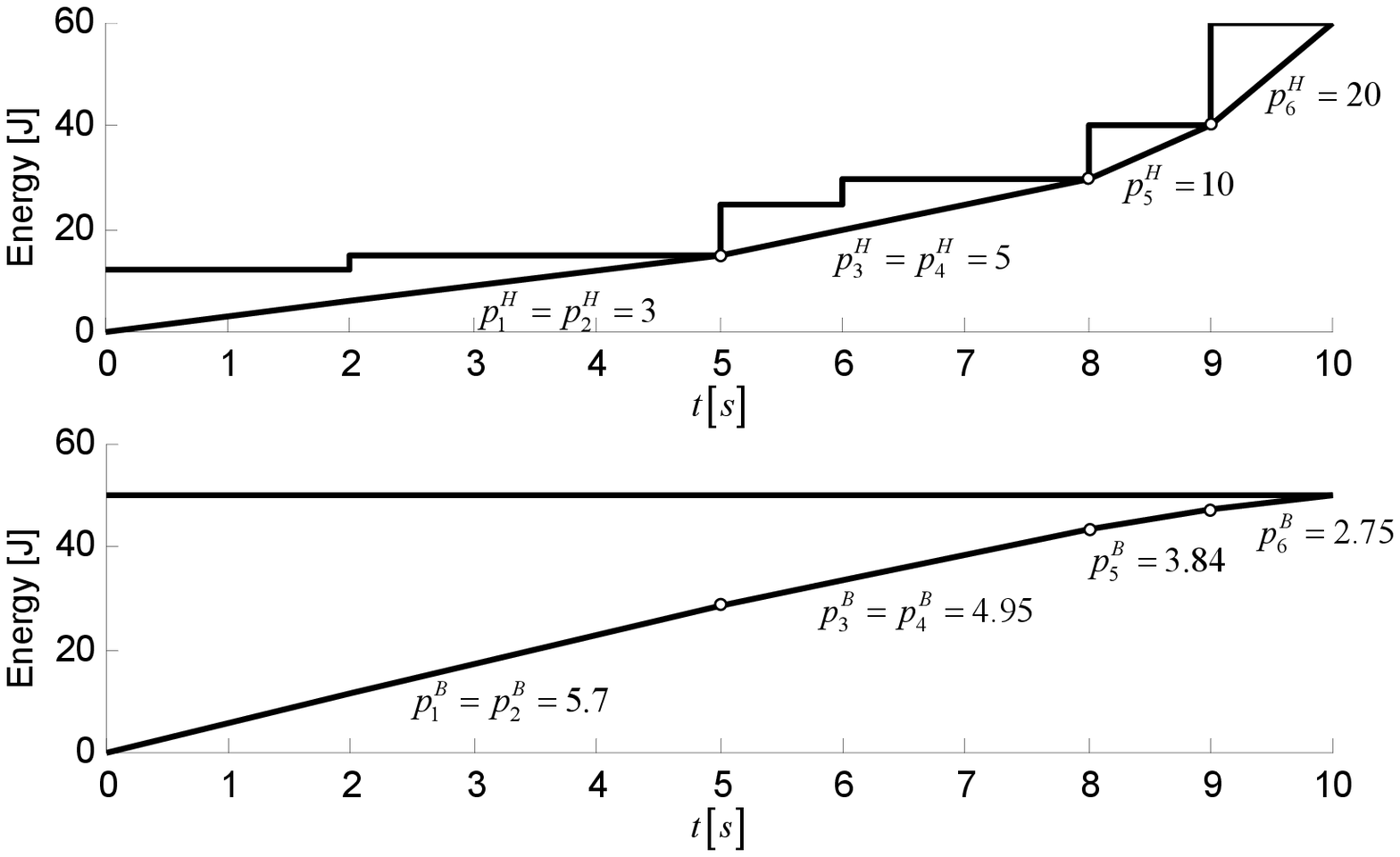}
      \vspace{-0.2cm}
   \caption{Optimal power allocation: EH and BO sensors ($T = 10$, $E_0^2 = 50 J$)}
   \vspace{-0.3cm}
   \label{fig.Result_1}
\end{figure}
Finally, Fig. \ref{fig.Result_1} depicts the optimal transmission policies
corresponding to the EH and BO sensors for a specific realization of
the energy arrivals. Clearly, (i) it satisfies all the lemmas and
theorems; (ii) Lemma \ref{Lemma4} on the monotonicity of the optimal
power allocation does not hold for the BO sensor; and (iii) in order
to collaboratively transmit data, the BO sensor must adopt an
optimal transmission policy which is different from that of the
single-sensor scenario, that is, constant transmit power within
$[0\dots T]$.
\subsection{{Computational complexity analysis}}
To recall, the computation of the optimal transmission policy for
the EH sensor entails the determination of a number of piece-wise
linear functions with minimal slope which connect a \emph{subset} of
the $N$ corner points on the cEH curve (see proof of Theorem
{\ref{Th:policy_energy_harvester}} and {\cite{Yang_Ulukus_2011})}.
In the worst case, the total number of corner points on the EC curve
equals{\footnote{The actual number depends on the specific
realization of energy arrivals.}} $N$. For the first corner point
(actually, the origin), the total number of slopes to be checked
equals $N$, that is, as many as the number of corner points up to
$t=T$. For the second corner point, the total number of slopes
equals $N-1$. The total number of operations is, thus,
{$N+(N-1)+\ldots 1=\frac{N\cdot (N-1)}{2}$}. Hence, the complexity
associated to the computation of the optimal transmission policy for
the EH sensor is {$\mathcal{O}(N^2)$}. As for the BO sensor, each
iteration of Algorithm {\ref{alg_battery}} entails the computation
of $N$ transmit powers (Steps 6 to 10). When a bi-section scheme is
adopted (rather than the grid search we actually used in Algorithm
{\ref{alg_battery})}, the total number of iterations needed is on
the order of {$\log(\frac{1}{\epsilon})$}
{\cite{Burden_Numerical_Analysis}}, where $\epsilon$ denotes the
constraints prescribed tolerance. Hence, the complexity associated
to the computations of the optimal transmission policy for the BO
sensor is {$\mathcal{O}(N \log(\frac{1}{\epsilon}))$}. In
conclusion, the computational complexity of the proposed scheme is
dominated by that of the algorithm presented in
{\cite{Yang_Ulukus_2011}} and it reads {$\mathcal{O}(N^2)$}.
%
%

%
\vspace{-0.3cm}
\section{Generalization to scenarios with finite storage capacity}
\vspace{-0.1cm}
\label{sec.Finite_capacity} Unlike in previous sections, here we
assume that the energy storage capacity of the EH sensor,
$E_{\text{max}}$, is \emph{finite}. If, in the $k$-th event, the
energy harvested by the EH sensor $E_{k}^{1}$ exceeds the remaining
storage capacity at that time instant, a \emph{battery overflow}
occurs. That is, its re-chargeable battery gets fully charged and
the excess harvested energy is simply discarded. In Appendix C, we
prove that any transmission policy allowing battery overflows to
occur is strictly suboptimal. Assuming that\footnote{Otherwise, part
of the energy in each arrival will be unavoidably wasted.}
$E_{k}^{1}\leq E_{\text{max}}$ $\forall k$, those suboptimal
solutions can be removed from the feasible set by imposing that
\vspace{-0.1cm}
\begin{equation}
e^H\!(t) \geq E^S\!(t) \triangleq  \sum_{k:s_k < t}
E_k^1-E_{\text{max}}
\label{eq.power_allocation_battery_capacity_constraint}
\end{equation}
for $0 \leq t \leq T$, where $E^S\!(t)$ denotes the cumulative
energy \emph{storage} (cES) constraint. One can easily verify that
Lemmas \ref{Lemma2}-\ref{Lemma3} and Lemma \ref{Lemma5} still hold
for the case of finite storage capacity. On the contrary, Lemma
\ref{Lemma4} does not, as we will discuss in the proof of Lemma
\ref{Lemma7}. Since, in particular, Lemma \ref{Lemma2} does hold,
the optimization problem can be posed again by the set of equations
given by \eqref{eq:score_function}-\eqref{eq.optimization_problem_1}
in Section \ref{sec.OptimalPowerAllocationPolicy}, along with the
additional constraint
\eqref{eq.power_allocation_battery_capacity_constraint}, namely,
\vspace{-0.1cm}
\begin{equation}
\sum_{k=1}^n \tau_k p_k^H \geq E_n^S = \sum_{k=0}^{n} E_k^1 -
E_{\text{max}} \; \text{for}\; n=1\ldots N.
\label{eq.battery_constraint_1}
\end{equation}
\begin{figure}[t]
   \centering
   \includegraphics[width=0.99\columnwidth]{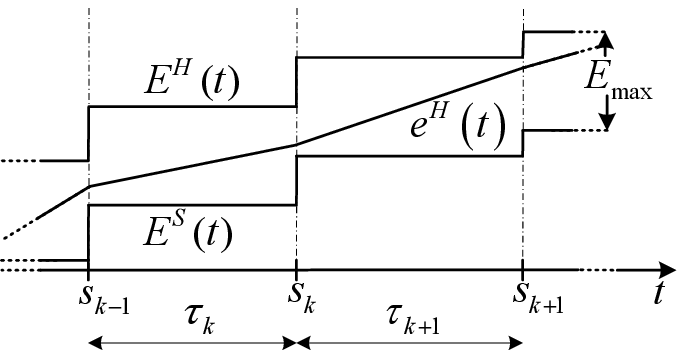}
      \vspace{-0.2cm}
   \caption{Cumulative energy harvesting and energy storage constraints.}
   \vspace{-0.3cm}
   \label{fig.Battery_capacity}
\end{figure}
A graphical representation for this additional constraint can be
found in Fig. \ref{fig.Battery_capacity}. Clearly, a transmit policy
is now feasible when the corresponding EC curve lies inside the
\emph{tunnel} defined by the cEH and cES curves. The additional
constraint \eqref{eq.battery_constraint_1} is affine and therefore
the optimization problem continues to be convex.

%
The next Lemma is an extension of Lemma \ref{Lemma6} for the case of
finite storage capacity:
\begin{Lem1}
\label{Lemma7} The jointly optimal power allocation policy when the
storage capacity of the EH sensor is finite is such that, whenever
the transmit power changes, its re-chargeable battery is either
fully charged or completely depleted.
\end{Lem1}
\begin{figure}[t]
   \centering
   \includegraphics[width=0.99\columnwidth]{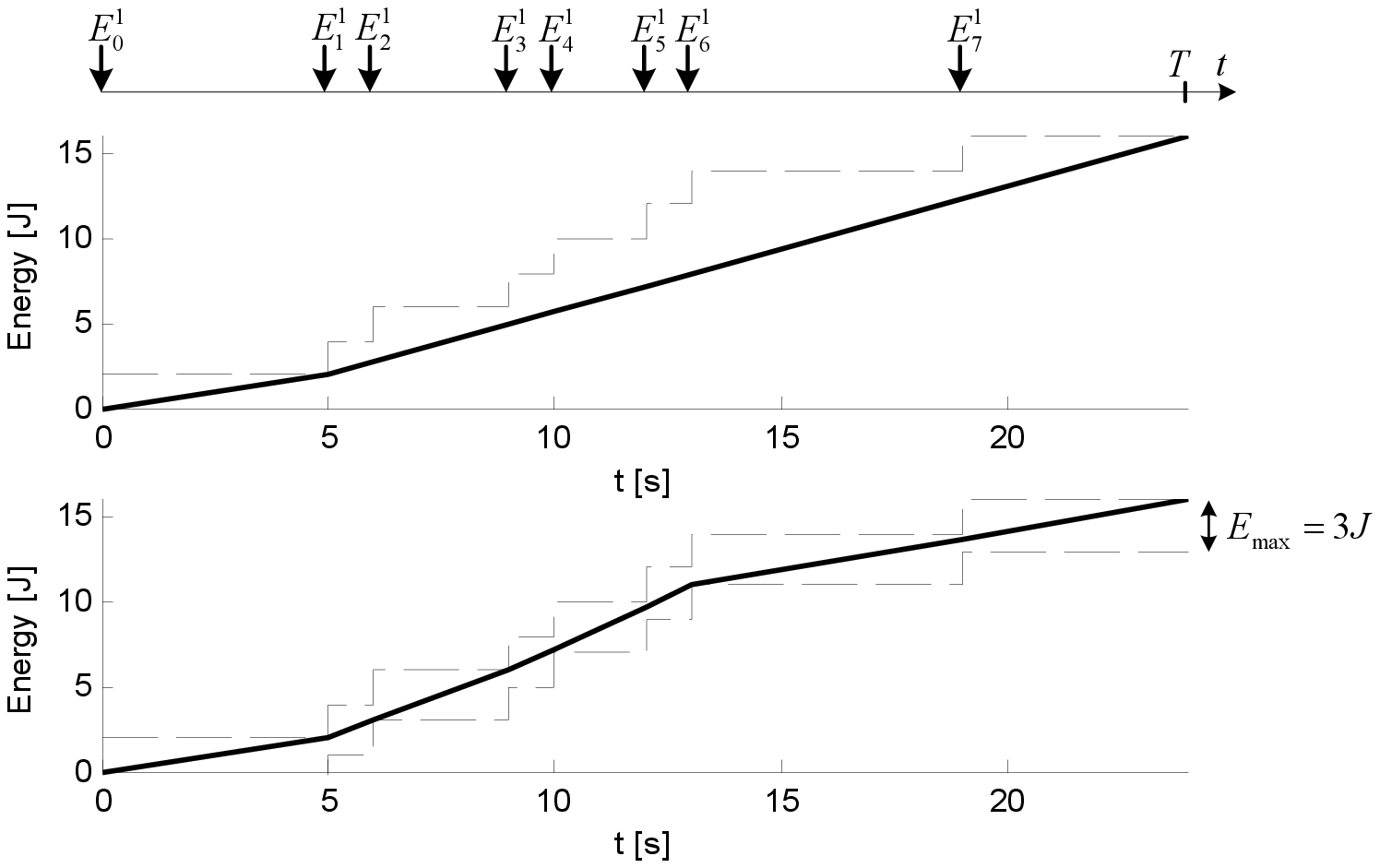}
      \vspace{-0.2cm}
   \caption{Optimal transmission policy for the EH sensor with infinite (top) and finite (bottom) battery capacity.}
   \vspace{-0.3cm}
   \label{fig.Battery_capacity_finite_infinite}
\end{figure}
\begin{proof}[Proof]
The proof of this Lemma is based again on the Karush-Kuhn-Tucker
(K.K.T.) conditions associated to the new optimization problem and
it can be found in Appendix D. This Lemma can also be regarded as an
extension of Lemma 4 in \cite{Tutuncuoglu_Yener_2012} for scenarios
with \emph{multiple} sensor nodes forming a virtual array. In
essence, Lemma \ref{Lemma7}  states that changes in the slope of the
EC curve can only occur when it hits either the cEH curve (depleted
battery) or the cES curve (fully charged). Intuitively, this is the
reason why Lemma \ref{Lemma4} (on the monotonically increasing
behavior of transmit powers for the EH sensor) does not hold anymore
in scenarios with finite energy storage capacity. This extent is
illustrated in Fig. \ref{fig.Battery_capacity_finite_infinite}.
\end{proof}
In the same vein of Theorem \ref{Th:policy_energy_harvester}, one
can easily verify that Lemmas \ref{Lemma2}-\ref{Lemma3},
\ref{Lemma5}, and \ref{Lemma7} unequivocally determine the optimal
transmission policy for the EH harvesting sensor with finite storage
capacity. Since those Lemmas are equivalent to the ones presented in
\cite{Tutuncuoglu_Yener_2012} for the \emph{single} sensor case, the
(jointly) optimal transmission policy for the EH sensor here can be
again \emph{independently} computed on the basis of algorithm A1
proposed therein. Interestingly, the optimal EC curve turns out to
be the shortest \emph{feasible} string which, now, lies
\emph{inside} the tunnel given by the cEH and cES curves. Besides,
the equation in Step 8 of Algorithm \ref{alg_battery} in our paper
continues to provide a connection between the primal and dual
solutions of the problem with finite storage capacity (yet with a
different definition of variables $A_k$, see Appendix D). Since no
additional constraints apply to the BO sensor, its optimal
transmission policy can be computed from that of the EH one with
Algorithm 1.
\vspace{-0.3cm}
\section{Computer Simulation Results}
\vspace{-0.1cm}
In this section, we assess the performance of the proposed power
allocation algorithm in a scenario where solar energy is harvested
from the environment. The energy storage system in the EH sensor
comprises (i) a supercapacitor \cite{Saggini_supercap_storage}; and
(ii) a re-chargeable Lithium-Ion battery. Upon being harvested, the
energy is temporarily stored in the supercapacitor. When it is fully
charged, the energy is transferred to the battery in a
burst\footnote{Pulse charging is beneficial for Lithium-Ion
batteries in terms of improved discharge capacity and longer life
cycles\cite{Li_pulse_charging_lion_batteries}.}. Clearly, this
validates the event-based model of the energy harvesting process
presented in Section \ref{sec:signal_model}. For such devices, the
amount of energy harvested in each event is constant and it equals
the maximum energy that can be stored in the supercapacitor. Since
solar irradiation levels change over time (e.g. from dawn to noon,
from winter to summer), so does the \emph{average} number of energy
arrivals (events). Consequently, the stochastic process that models
energy arrivals is non-stationary. In the sequel, we adopt a Poisson
process with time-varying mean given by $\lambda_{E}(t)$. From the
solar irradiation data in \cite{NSRDB}, the mean arrival rate from 5
A.M. to 12 P.M. (i.e. dawn to noon, with $T=7$ h) can be fitted by
the following exponential function:
\vspace{-0.1cm}
\begin{equation}
\label{eq.solar_data} \lambda_E(t) = \beta_{E, c,T} e^{ct}
\end{equation}
\begin{figure}[t]
   \centering
   \includegraphics[width=0.99\columnwidth]{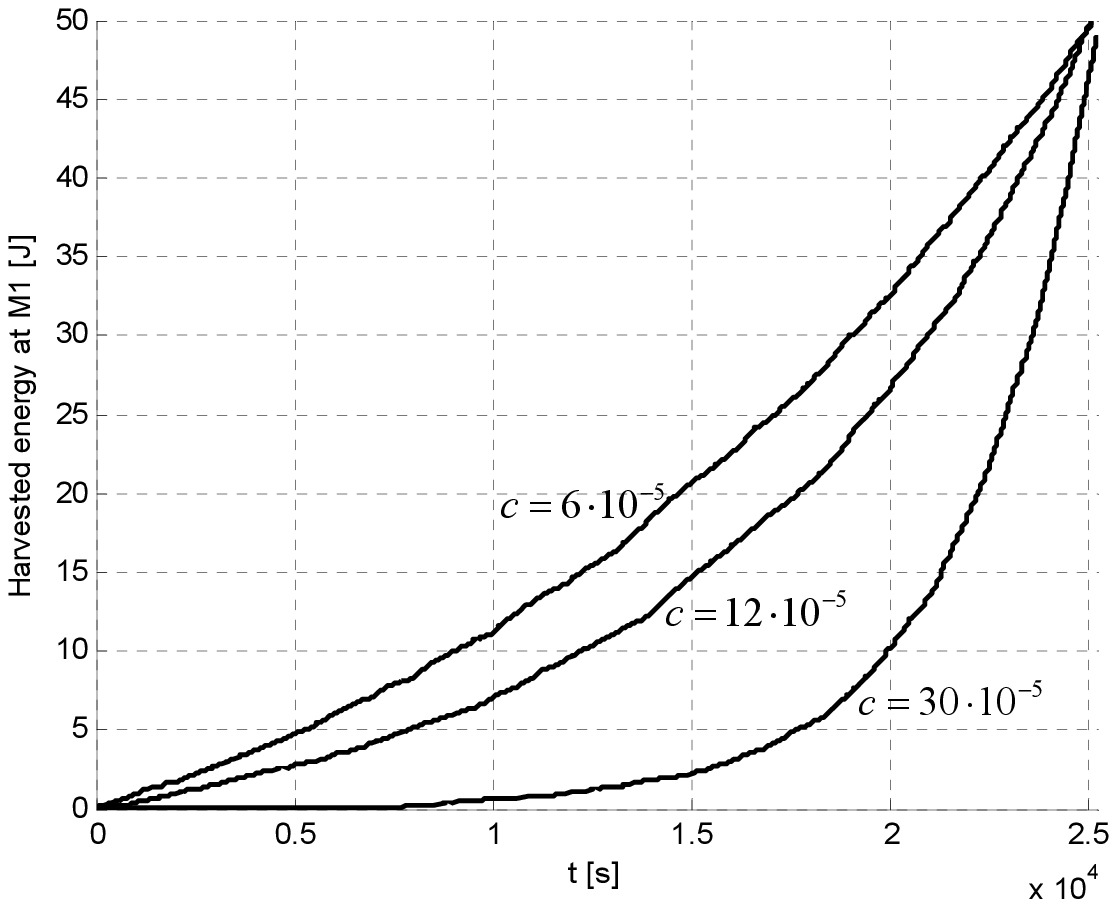}
      \vspace{-0.2cm}
   \caption{Typical cumulative energy curves parameterized by $c$.}
   \vspace{-0.3cm}
   \label{fig.Result_5}
\end{figure}
where parameter $c$ models the variability of the energy harvested
over time (i.e. the rate of energy transfers from the capacitor to
the battery); and $\beta_{E, c,T}$ is a constant depending on the
total amount of energy harvested $E$, parameter $c$, and the total
harvesting time $T$. For the solar irradiation data in \cite{NSRDB},
it yields $\beta_{E, c,T}=3.899\cdot 10^{-2}$ and $c=6\cdot
10^{-5}$. Figure \ref{fig.Result_5} shows a number of cumulative
energy harvesting curves for different values of parameter $c$: the
higher $c$, the higher the energy variability, i.e. the steeper the
curves by the end of the observation interval. Hereinafter, we let
$E_{T}^H=\sum_{k=0}^{N-1} E_k^1$ and $E_{T}^B=E_0^2$ denote the
total energy harvested by/stored in the EH and BO sensors,
respectively; whereas $E_{T}=E_{T}^H+ E_{T}^B$ accounts for the
total energy in the system. Further, we define
$R_E={E_{T}^B}/{E_{T}^H}$ as the ratio between the total energy in
the BO and EH sensors, that is, for large $R_E$, the battery
operated sensor dominates. In all plots, we have $T=7$ h (from 5
A.M. to noon). \vspace{-0.3cm}
\subsection{Infinite Energy Storage Capacity}
\vspace{-0.1cm}
\begin{figure}[t]
   \centering
   \includegraphics[width=0.99\columnwidth]{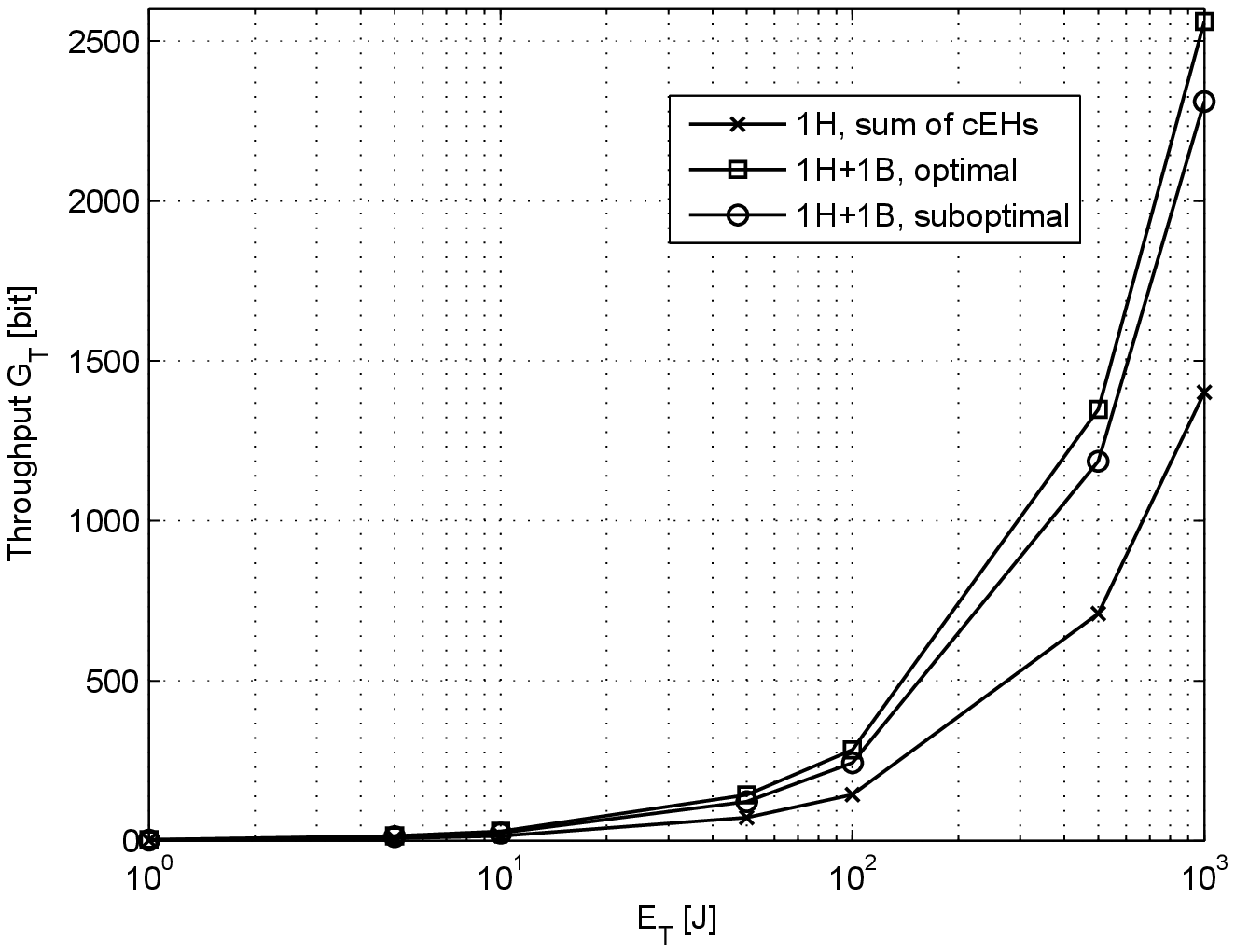}
      \vspace{-0.2cm}
   \caption{Throughput vs. total energy in the system ($c = 30 \cdot 10^{-5}$, $R_E = 1$). }
   \vspace{-0.3cm}
   \label{fig.Result_2}
\end{figure}
Initially, we assume that the storage capacity of the aforementioned
Lithium-ion battery is infinite. In Fig. \ref{fig.Result_2}, we
depict the throughput of the virtual array with the jointly optimal
transmission policies defined by Theorem
\ref{Th:policy_energy_harvester} and Theorem
\ref{Th:policy_battery_operated} for the EH and BO sensors,
respectively. The amount of energy harvested by/stored in the EH and
BO sensors is identical ($R_E=1$) and results are shown as a
function of the total energy $E_T$ in the system. As benchmarks, we
consider (i) a system with only one EH sensor, the cEH curve of
which is given by the point-wise sum of the cEH curves for the EH
and BO sensors in the virtual array (curve labeled with ``1H, sum of
cEHs''); and (ii) a two-sensor virtual array in which the
transmission policies for the EH and BO sensors are optimized
individually for each sensor as in \cite{Yang_Ulukus_2011}, which is
suboptimal for a virtual array (``1H+1B, suboptimal''). For (ii),
the optimal policy for the BO sensor consists in a constant transmit
power for $t=0\ldots T$. Unsurprisingly, for systems with multiple
transmitters the beamforming gain translates into substantially
higher throughputs. Besides, some additional throughput gain results
from the \emph{joint} optimization of the transmission policies for
the EH and BO sensors, that is, by forcing the BO sensor to adapt to
the changes in transmit power in the EH sensor.

\begin{figure}[t]
   \centering
   \includegraphics[width=0.99\columnwidth]{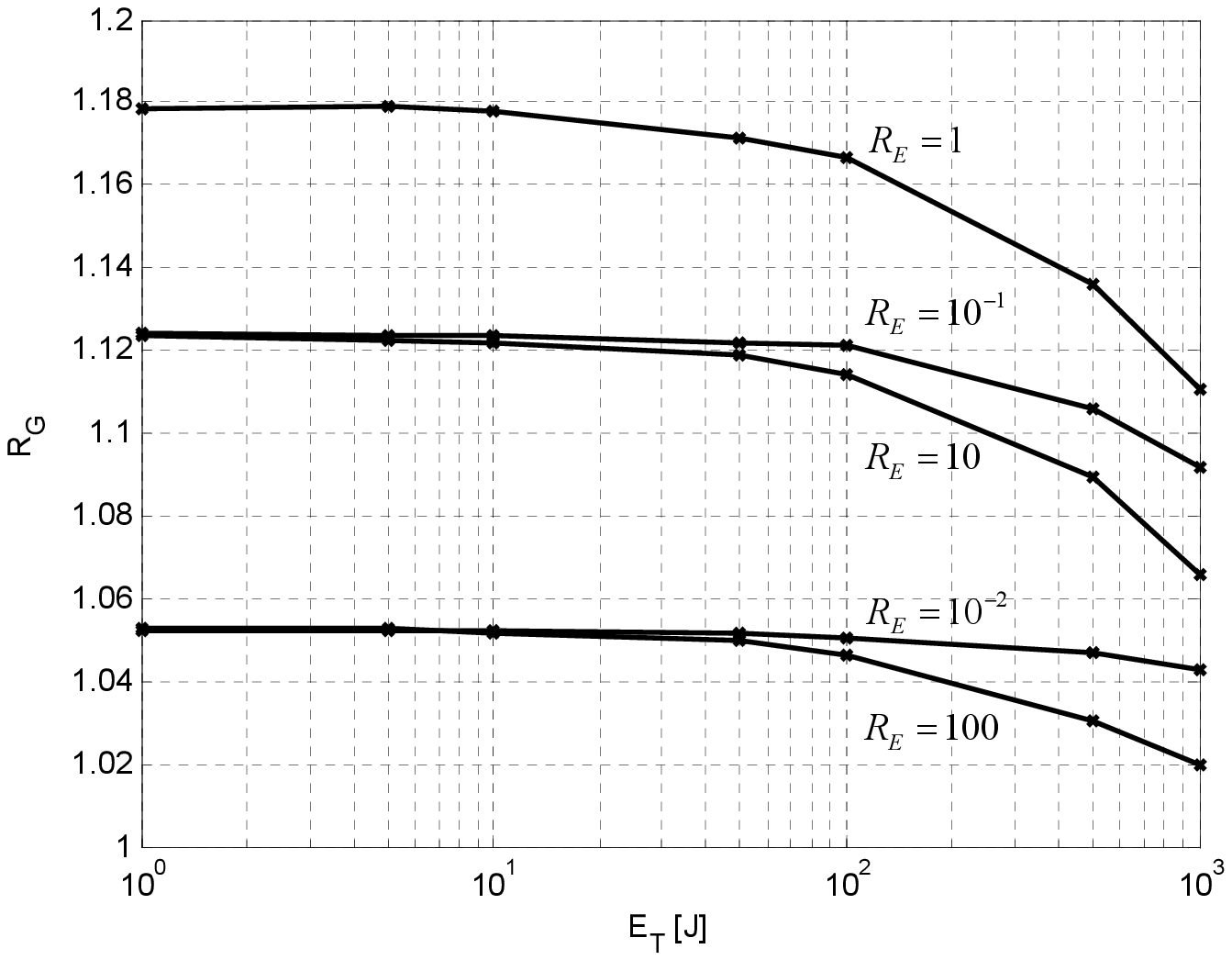}
      \vspace{-0.2cm}
   \caption{Throughput gain vs. total energy in the system ($c= 30 \cdot 10^{-5}$).}
   \vspace{-0.3cm}
   \label{fig.Result_3}
\end{figure}
Fig. \ref{fig.Result_3} provides further insights on the throughput
gains stemming from the \emph{joint} optimization of transmission
policies over sensors. More precisely, we depict the {throughput
gain by} ratio $R_G = {G_{T,\text{opt}}}/{G_{T,\text{subopt}}}$ as a
function of the total system energy. Interestingly, the highest gain
is attained when the total energy harvested by the EH sensor equals
that stored in the BO one, that is, for $R_E=1$. Yet in a totally
different context, this is consistent with \cite{Pun2009} where the
authors conclude that, in order to maximize the beamforming gain,
the received signal levels from the opportunistically selected
sensors must be comparable. Conversely, when either the EH or the BO
sensors dominate ($R_E\ll 1$ or $R_E\gg 1$, respectively) the gain
from a joint optimization becomes marginal ($R_G\rightarrow 1$)
since the signal received from the other sensor is weak. We also
observe that, in the case of unbalanced energy levels, throughput
gains are lower when the BO sensor dominates. This is motivated by
the fact that when $E_T^H\ll E_T^B$ the policy for the EH sensor has
very little impact in the definition of the (jointly) optimal policy
for the BO one. In other words, the energy consumption curves for
the BO sensor with and without joint optimization are similar and,
hence, the {throughput gain} approaches 1. It is also clear that
throughput gains become negligible when $E_T$ increases (i.e. in the
high SNR regime). Let $\alpha = E_{T,\text{high}}/E_{T,\text{low}}$
denote the ratio of total system energies in the high and low SNR
regimes. Since the total received power $p_{BF}\!(t)$ scales with
$\alpha$, from the score function in \eqref{eq:score_function} and
for large $E_{T,\text{high}}$ we can write, $$R_G
(E_{T,\text{high}})= \frac{N\log (\alpha) +
G_{T,\text{opt}}(E_{T,\text{low}})}{N\log (\alpha) +
G_{T,\text{subopt}}(E_{T,\text{low}})}.$$ Clearly, for large
$\alpha$ the impact of the specific transmission policies (optimal/
suboptimal) diminishes. In other words, joint optimization of
transmission policies is more relevant in the low-SNR regime.

\begin{figure}[t]
   \centering
   \includegraphics[width=0.99\columnwidth]{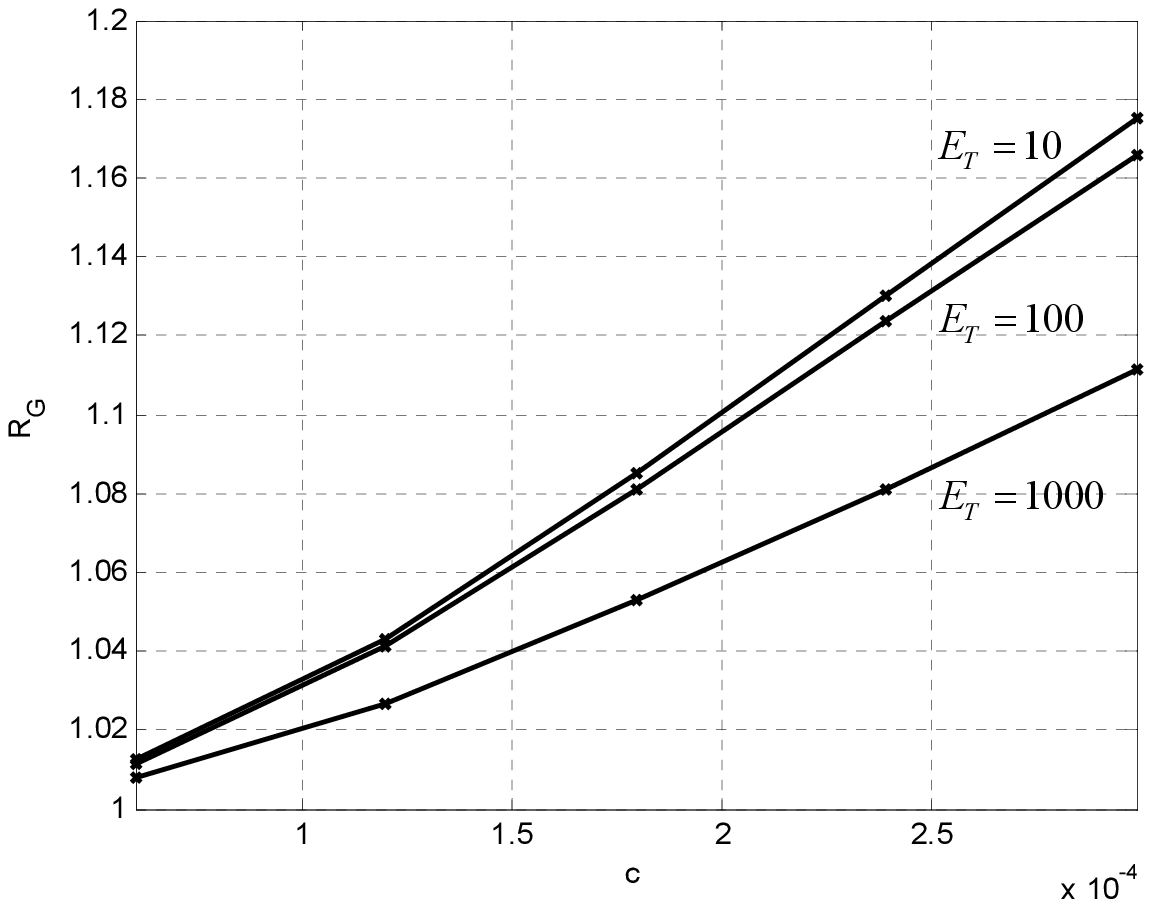}
      \vspace{-0.2cm}
   \caption{Throughput gain vs. variability of the energy harvested ($R_E = 1$).}
   \vspace{-0.3cm}
   \label{fig.Result_4}
\end{figure}
Next, Fig. \ref{fig.Result_4} illustrates the impact of the
variability of energy arrivals in the throughput gain. Clearly, the
higher the variability (i.e. for higher values of parameter $c$),
the higher the gain: $R_G=1.2$ (or $+20\%$ gain) for $c=3\cdot
10^{-4}$ and $E_T=10$ J. On the contrary, if the average number of
arrivals does not vary (increase) substantially in the observation
interval, the gain stemming from a joint optimization of both
transmission policies is marginal ($R_G\approx 1$). In conclusion,
rapid variations of solar irradiation levels from dawn to noon (e.g.
in high latitude locations, winter time) make joint optimization of
transmission policies advisable.\\
\vspace{-0.3cm}
\subsection{Finite Energy Storage Capacity}
\vspace{-0.1cm}
Unlike in the previous subsection, here we realistically assume that
the energy storage capacity for the EH sensor is finite.

\begin{figure}[t]
   \centering
   \includegraphics[width=0.99\columnwidth]{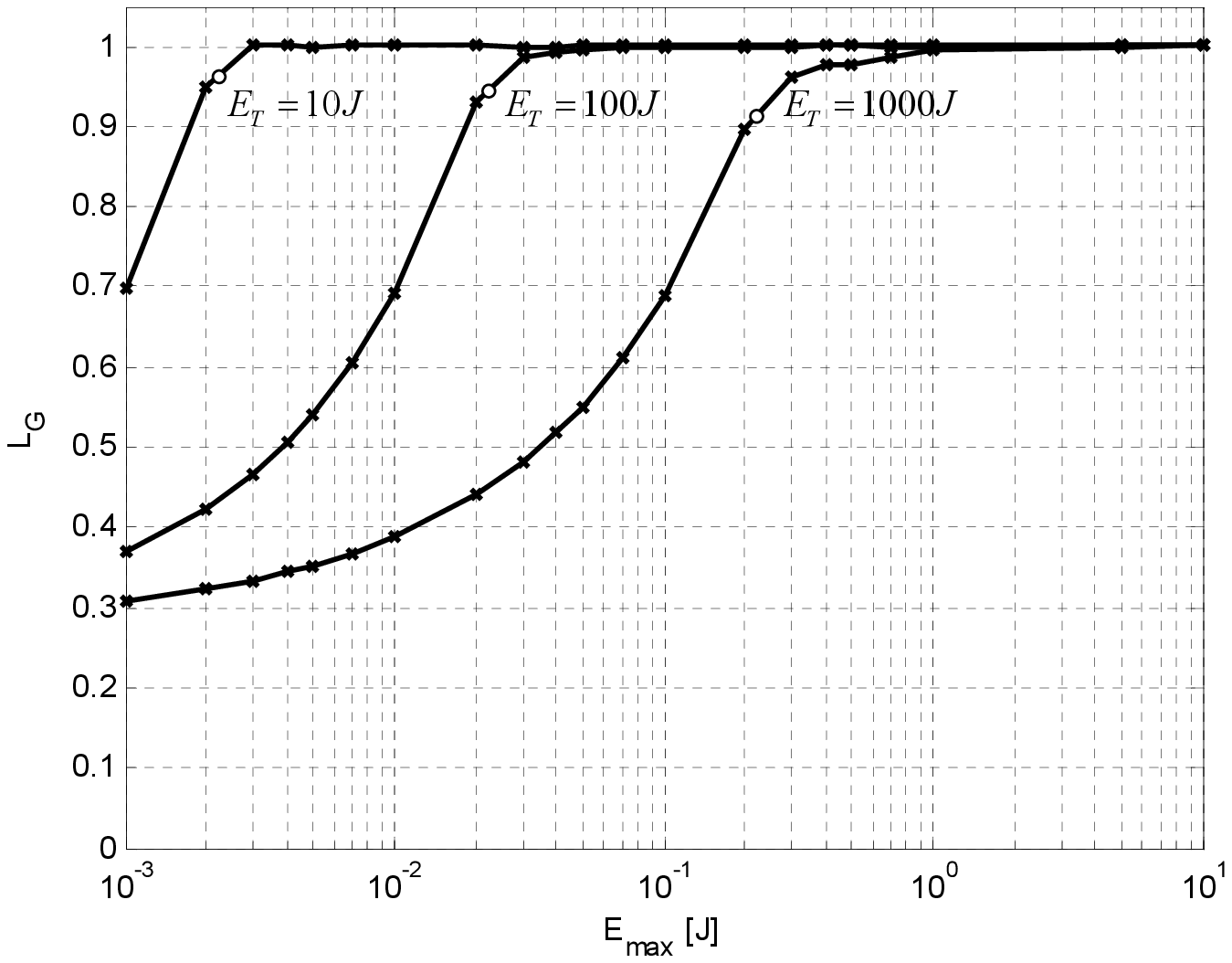}
      \vspace{-0.2cm}
   \caption{Throughput {ratio (loss)} as a function of battery capacity ($N=2250$ epochs). Big round markers on the curve correspond to the operating points where the maximum storage capacity $E_{\text{max}}$ equals the energy harvested in each arrival $E_k^1$ ($E_k^1=2.21\cdot 10^{-3},2.21\cdot 10^{-2}, 2.21\cdot 10^{-1}$ for the $E_T=10,100,1000$ J curves, respectively).}
   \vspace{-0.3cm}
   \label{fig.thr_loss_capacity_finite_infinite}
\end{figure}
Figure \ref{fig.thr_loss_capacity_finite_infinite} depicts the total
loss in throughput with respect to the case of infinite storage
capacity {by throughpout ratio $L_G$}. Interestingly, as long as the
maximum storage capacity is greater than the energy harvested in
each arrival, the throughput loss is barely noticeable {(the
throughput ratio equals 1)}. In other words, the changes in the
optimal transmission policy resulting from the introduction of the
additional constraint \eqref{eq.battery_constraint_1}, which avoids
battery overflows, have a rather marginal impact on the achievable
throughput. This is excellent news since, typically, storage
capacity is well above individual harvested energy levels. On the
contrary, throughput performance rapidly degrades for smaller
storage capacities. This stems from the fact that now part of the
energy in each arrival is unavoidably wasted in battery overflows.
As a result, the total amount of energy stored with respect to the
case of infinite capacity decreases, and so does the resulting
throughput.

\begin{figure}[t]
   \centering
   \includegraphics[width=0.99\columnwidth]{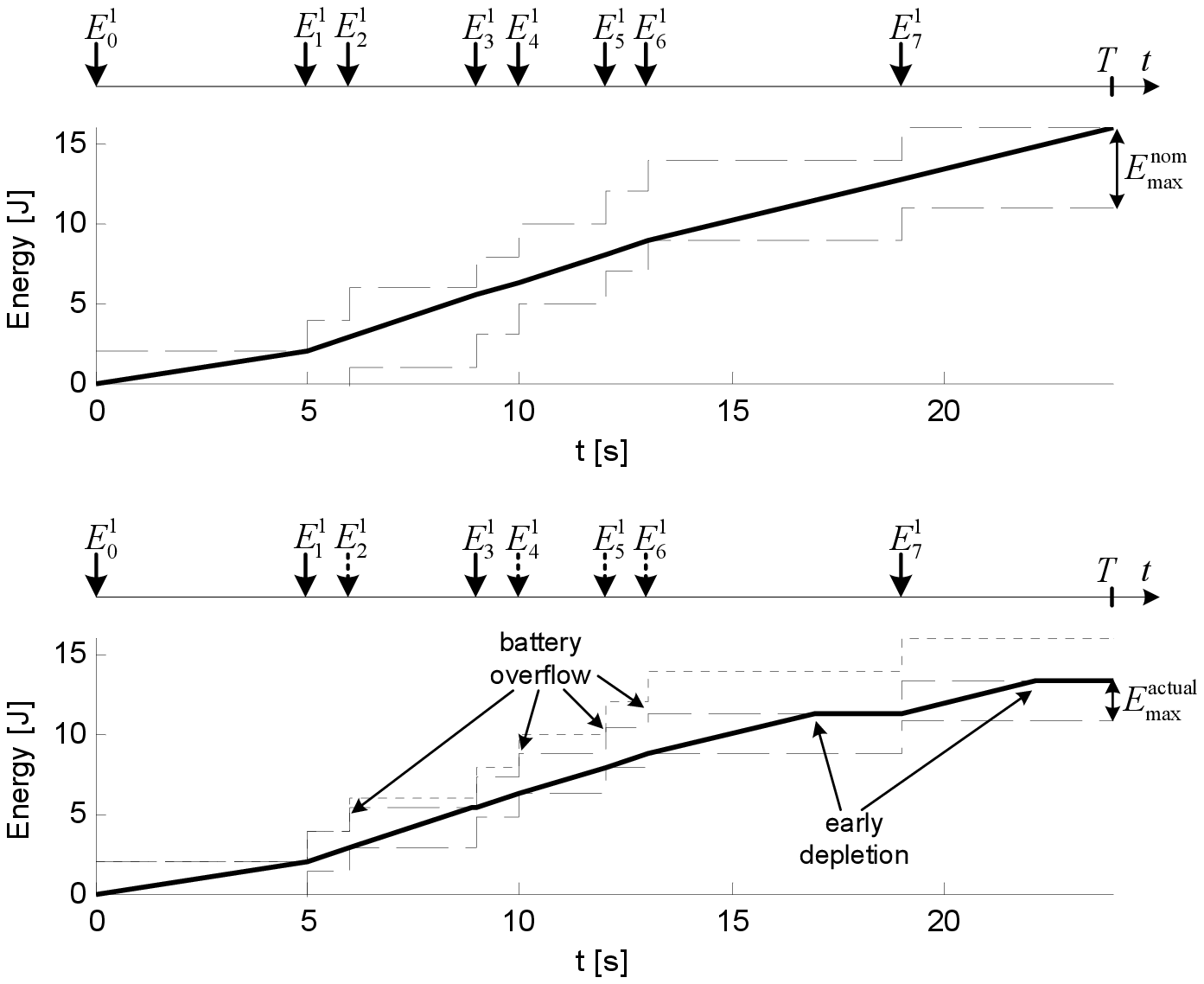}
      \vspace{-0.2cm}
   \caption{Battery overflow and early depletion phenomena: transmission policies for the EH sensor with nominal battery capacity (top, $E_{\text{max}}^{\text{nom}}=5$ J) and actual capacity (bottom, $E_{\text{max}}^{\text{actual}}=2.5$ J) for a given realization of energy arrivals. Dotted arrows indicate the arrivals in which part of the energy is wasted ($E_k^1 = 2$ J). As a
   reference, the lower plot includes the cEH curve for the nominal
   capacity (dash-dotted line).}
   \vspace{-0.3cm}
   \label{fig.Battery_capacity_suboptimal_overflow}
\end{figure}
Next, we analyze the impact of battery degradation in the EH sensor
on system performance. Our focus is on impairments due to
\emph{long-term} battery degradation due to e.g. aging. Accordingly,
its storage capacity is assumed to take a constant value for the
whole transmission period (i.e. no battery leakage between
arrivals). The \emph{nominal} storage capacity
$E_{\text{max}}^{\text{nom}}$, on which basis the optimal
transmission policies for the EH and BO are computed, is assumed to
be known. On the contrary, the \emph{actual} capacity
$E_{\text{max}}^{\text{actual}}\leq E_{\text{max}}^{\text{nom}}$,
which enables data transmission, is unknown. The fact that the
actual capacity is lower that its nominal value may result into
\emph{battery overflows} and \emph{early battery depletion} (see
Fig. \ref{fig.Battery_capacity_suboptimal_overflow}), both having a
negative impact on the achievable throughput. Despite of the
introduction of the additional constraint
\eqref{eq.battery_constraint_1}, now there is a risk to waste part
of the energy arrivals in battery overflows since the remaining
battery capacity is smaller than expected. As an example, for the
particular realization in Fig.
\ref{fig.Battery_capacity_suboptimal_overflow}, the total energy
actually harvested within $0\ldots T$ amounts to $13.375$ J instead
of $16$ J. Likewise, the fact that the actual energy stored in the
battery is lower than expected might lead to early battery
depletions. This forces data transmission for the EH sensor to be
suspended until the next energy arrival. Consequently, the
beamforming gain vanishes for this period of time.

\begin{figure}[t]
   \centering
   \includegraphics[width=0.99\columnwidth]{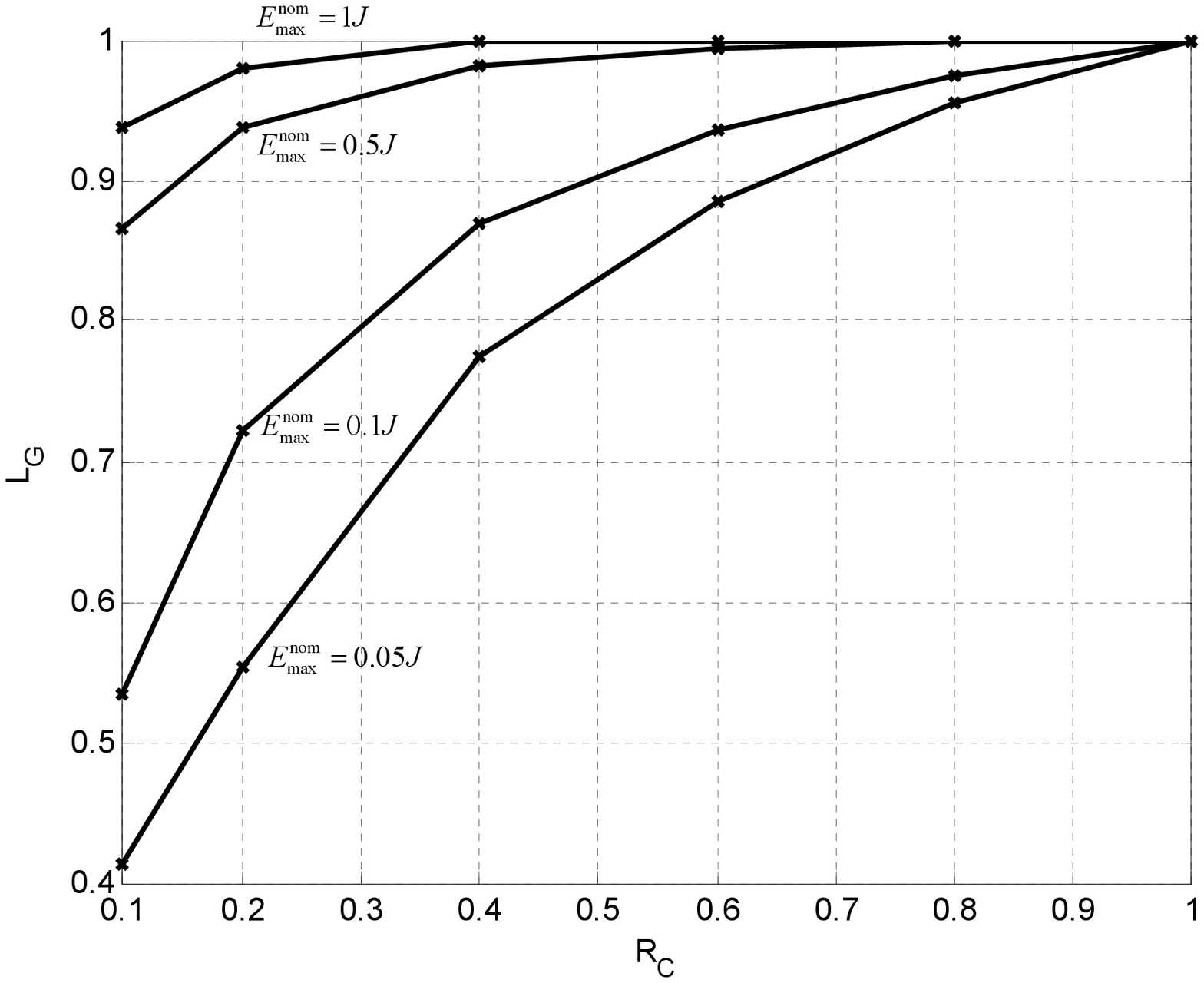}
      \vspace{-0.2cm}
   \caption{Throughput {ratio (loss)} due to battery capacity degradation ($E_T = 100$ J, $c = 30 \cdot 10^{-5}$, $T = 7h$)}
   \vspace{-0.3cm}
   \label{fig.Result_battery_capacity}
\end{figure}
In Fig. \ref{fig.Result_battery_capacity}, we investigate the impact
of battery overflows and early depletions on throughput. More
specifically, we depict {the throughput ratio $L_G =
{G_{T,\text{actual}}}/{G_{T,\text{nom}}}$} as a function of the
ratio between actual and nominal battery capacities, namely, $R_C =
{E_{\text{max}}^{\text{actual}}}/{E_{\text{max}}^{\text{nom}}}$.
Unsurprisingly, throughput degradation is particulary severe and
faster for smaller values of nominal capacity (i.e. for
$E_{\text{max}}^{\text{nom}}=0.05$ J). In this case, the amount of
energy in each arrival ($E_k^1 = 2.21 \cdot 10^{-2}$ J) is
comparable to the nominal capacity. Consequently, many battery
overflows and early depletions occur. Furthermore, for $R_C = 0.1$,
the actual battery capacity amounts to
$E_{\text{max}}^{\text{actual}}=5 \cdot 10^{-3}$ which is below
$E_k^1$. Hence, every energy arrival causes a battery overflow which
results into a throughput loss of $60\%$. It is also worth noting
that for large nominal battery capacities
($E_{\text{max}}^{\text{nom}}=1$ J) and higher values of capacity
degradation ($R_C = 0.1$) there is also a noticeable throughput loss
(some $10\%$). Even though the actual battery capacity
{($E_{\text{max}}^{\text{actual}}=0.1$ J)} is well above $E_k^1$,
the mismatch between nominal and actual capacities results into some
battery overflows and early depletions too.
\vspace{-0.3cm}
\section{Conclusions}
\vspace{-0.1cm}
In this paper, we have derived the \emph{jointly} optimal
transmission policy which allows an energy harvesting plus a battery
operated sensor node to act as a virtual antenna array to maximize
throughput for a given deadline. The necessary conditions for
optimality that we have identified, both for scenarios with infinite
and finite energy storage capacity in the energy harvesting sensor,
allowed us to prove that the optimal transmission policy for the
energy harvesting sensor can be computed independently from that of
the battery operated one according to the procedure described in
\cite{Yang_Ulukus_2011} and \cite{Tutuncuoglu_Yener_2012},
respectively. Interestingly enough, we have proved that such
policies \emph{continue} to be optimal for our \emph{two}-sensor
(vs. single-sensor) scenario. Moreover, we have shown that the
optimal transmission policy for the battery operated sensor is
unequivocally determined and can be iteratively computed from that
of the energy harvesting one. The resulting policy is, in general,
different from that of battery operated sensors in single-sensor
scenarios (i.e. constant transmit power). The performance of the
jointly optimal transmission policy has been assessed by means of
computer simulations in a realistic scenario where solar energy is
harvested from the environment. Computer simulation results revealed
that, in scenarios with \emph{infinite} storage capacity in the
energy harvesting sensor, the joint optimization of transmit
policies in combination with beamforming yields substantial
throughput gains. The highest gain is attained when the total energy
in the energy harvesting and battery operated sensors are identical.
However, the gain becomes negligible in high-SNR scenarios where
large amounts of energy are harvested by/stored in sensors. In the
case of unbalanced energy levels, throughput gains are lower when
the BO sensor dominates. Besides, we have found that throughput gain
is larger when solar irradiation levels vary rapidly. We have also
learnt that throughput losses stemming from finite storage capacity
are only substantial when battery capacity is smaller than the
amount of energy in each arrival. Finally, we have observed that a
long-term degradation of battery capacity may result into battery
overflows and early battery depletions. The associated throughput
loss is particulary severe for smaller values of the nominal storage
capacity. Still, the impact of the mismatch between nominal and
actual capacities can also be noticeable for larger values.
\vspace{-0.3cm}
\section{Acknowledgements}
\vspace{-0.1cm} This work is partly supported by the project JUNTOS
(TEC2010-17816), NEWCOM\# (318306), Spanish Ministry of Education
(FPU grant AP2008-03952) and by the Catalan Government under 2009
SGR 1046.
%
\appendix
\vspace{-0.3cm}
\section{Proof of Lemma \ref{Lemma2}}
\vspace{-0.1cm}
\label{sec.Proof_Lemma2}
Assume that the optimal policy before $s_{k-1}$ and after $s_k$ is
optimal.  The total throughput in the $k$-th epoch is given by
$G_{\tau_k} = \int_{s_{k-1}}^{s_{k}} \log (1 + p_{BF}(t)) dt$ where,
to recall, we defined $p_{BF}\!(t) = (\sqrt{p^H\!(t)} +
\sqrt{p^B\!(t)})^2$ as the instantaneous power received at the base
station from the two sensors. Besides, let $\Delta e_{BF}=
\int_{s_{k-1}}^{s_{k}} p_{BF}(t) dt$ denote the total received
energy in the $k$-th epoch of duration $\tau_k=s_k-s_{k-1}$. From
Jensen's inequality \cite{Polyanin_Manizhirov_Math}[Sec. 7.2.5], we
have that the following inequality:
\vspace{-0.1cm}
\begin{equation}
\label{eq.Jensens_inequality} \frac{\int_a^b g\left(f(t)\right) h(t)
dt}{\int_a^b h(t) dt} \leq g \left(\frac{\int_a^b f(t) h(t)
dt}{\int_a^b h(t) dt} \right)  \\\nonumber
\end{equation}
holds as long as $g(\cdot)$ is a concave function, $f(t)$ is such
that $\alpha \leq f(t) \leq \beta$, and $h(t) \geq 0$. Letting $g(p)
= \log(1+p)$, $f(t) = p_{BF}(t)$ and $h(t) = 1$ yields
\vspace{-0.1cm}
\begin{align}
G_{\tau_k} &= \int_{s_{k-1}}^{s_{k}} \log (1 + p_{BF}(t)) dt \nonumber\\
\quad &\leq \tau_k \log\left( 1 + \left(\frac{\int_{s_{k-1}}^{s_{k}}
p_{BF}(t) dt}{\int_{s_{k-1}}^{s_{k}} dt} \right)
\right)\label{eq.Jensens_inequality_rate}\\
\quad &= \tau_k \log\left( 1 + \frac{\Delta e_{BF}}{\tau_k}
\right).\nonumber
\end{align}
This last inequality evidences that for a given energy $\Delta
e_{BF}$, the optimal power allocation policies for the $k$-th epoch
must be such that the instantaneous received power at the BS is
\emph{constant} and equal to $p_{BF}(t)=\Delta e_{BF}/\tau_k$. In
order to determine the optimal transmission policy for \emph{each}
sensor, we resort to Cauchy's inequality
\cite{Polyanin_Manizhirov_Math}[Sec. 7.2.5] to learn that
\vspace{-0.1cm}
\begin{equation}
\label{eq.Cauchy_inequality_power1} \left(\int_{s_{k-1}}^{s_{k}}
p_{BF}(t) dt\right)^{\frac{1}{2}} \leq \left(\int_{s_{k-1}}^{s_{k}}
p^H\!(t) dt\right)^{\frac{1}{2}} + \left(\int_{s_{k-1}}^{s_{k}}
p^B\!(t) dt\right)^{\frac{1}{2}}\nonumber
\end{equation}
\begin{figure}[t]
   \centering
   \includegraphics[width=0.99\columnwidth]{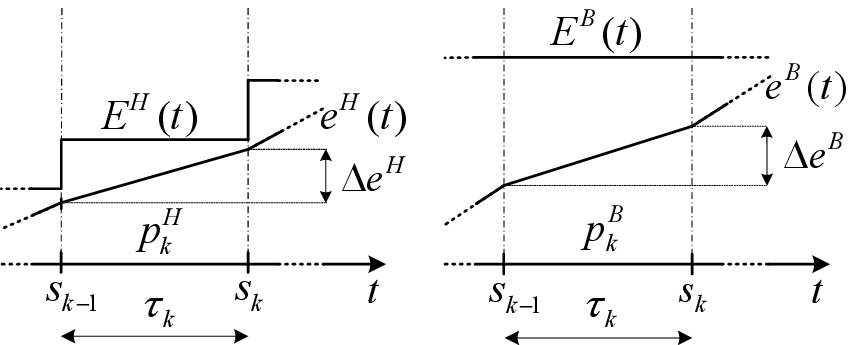}
      \vspace{-0.2cm}
   \caption{Transmit power in each sensor remains constant between consecutive events (Lemma \ref{Lemma2}).}
   \vspace{-0.3cm}
   \label{fig.uniform_power_property_2}
\end{figure}
or, equivalently (see Fig. \ref{fig.uniform_power_property_2}),
\vspace{-0.1cm}
\begin{equation}
\label{eq.Cauchy_inequality_power} \int_{s_{k-1}}^{s_{k}} p_{BF}(t)
dt \leq  \left(\sqrt{\Delta e^H} + \sqrt{\Delta e^B}\right)^2.
\end{equation}
By replacing \eqref{eq.Cauchy_inequality_power} into
\eqref{eq.Jensens_inequality_rate}, we finally get:
\vspace{-0.1cm}
\begin{equation}
\label{eq.Cauchy_inequality_power_final} G_{\tau_k}  \leq \tau_k
\log\left( 1 + \left(\sqrt{\frac{\Delta e^H}{\tau_k}} +
\sqrt{\frac{\Delta e^B}{\tau_k}}\right)^2  \right).
\end{equation}
In other words, the individual power allocation policies that
maximize the throughput in the $k$-th epoch consist in using a
\emph{constant} transmit power given by $p^H\!(t) = \Delta
e^H/\tau_k$ and $p^B\!(t) = \Delta e^B/\tau_k$ for the EH and BO
sensors, respectively. This concludes the proof. 
\vspace{-0.3cm}
\section{Proof of Lemma \ref{Lemma6}}
\begin{figure}[t]
   \centering
   \includegraphics[width=0.99\columnwidth]{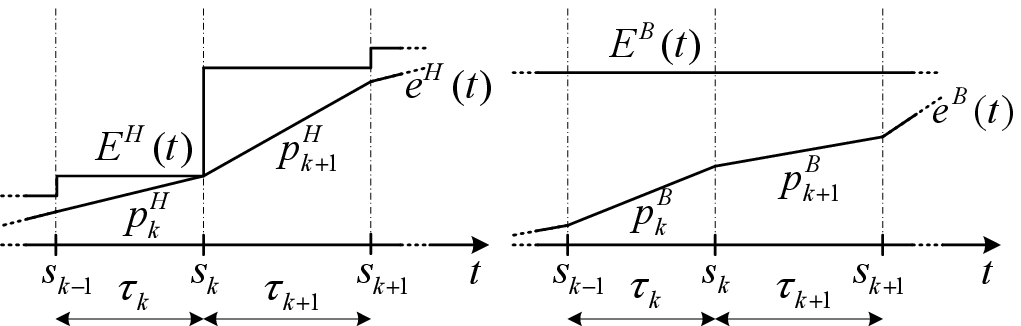}
      \vspace{-0.2cm}
   \caption{When transmit power changes, the energy consumed by the
EH sensor equals the energy harvested (Lemma \ref{Lemma6}).}
   \vspace{-0.3cm}
   \label{fig.non_uniform_power_property_2}
\end{figure}

\vspace{-0.1cm}
\label{sec:appendix_curve_touches} \label{sec.Proof_Lemma4} The
Lagrangian of the optimization problem
\eqref{eq.optimization_problem_1} is given by
\vspace{-0.1cm}
\begin{align}
\label{eq.optimization_problem_proof_lagrangian} \mathcal{L}_1 = &-
\sum_{k=1}^{N} \tau_k \log \left(1+(\sqrt{p^H_k} +
\sqrt{p^B_k})^2\right)\nonumber\\
\quad &+ \sum_{n=1}^{N} \lambda_n \left(\sum_{k=1}^{n} \tau_k p^H_k
- E_n^H\right)
+ \sum_{n=1}^{N} \nu_n \left(\sum_{k=1}^{n} \tau_k p^B_k - E_n^B\right)\nonumber\\
\quad &- \sum_{k=1}^{N} \mu_k p^H_k - \sum_{k=1}^{N} \xi_k p^B_k
\end{align}
and, hence, the corresponding K.K.T. conditions read %
\begin{align}
\frac{\partial
\mathcal{L}_1}{\partial p^H_k}, \frac{\partial \mathcal{L}_1}{\partial p^B_k} &= 0\label{eq_KKT_derivatives}\\
\sum_{k=1}^{n} \tau_k \breve{p}^H_k &\leq E_n^H \; \text{for}\;
n=1\ldots N \label{eq_KKT_causality_x}\\ \sum_{k=1}^{n} \tau_k
\breve{p}^B_k &\leq E_n^B \; \text{for}\; n=1\ldots N
\label{eq_KKT_causality_y}\\
\breve{p}^H_k,\breve{p}^B_k &> 0\label{eq_KKT_positivity_xy}\\
\breve{\lambda}_n, \breve{\nu}_n, \breve{\mu}_k, \breve{\xi}_k &\geq 0 \label{eq_KKT_positivity_lagrangian_mult}\\
\breve{\lambda}_n \left(\sum_{k=1}^{n} \tau_k \breve{p}^H_k - E_n^H\right) &= 0 \; \text{for}\; n=1\ldots N \label{eq_KKT_slackness_lambda}\\
\breve{\nu}_n \left(\sum_{k=1}^{n} \tau_k \breve{p}^B_k - E_n^B\right) &= 0 \; \text{for}\; n=1\ldots N \label{eq_KKT_slackness_nu}\\
-\breve{\mu}_k \breve{p}^H_k &= 0 \; \text{for}\; k=1\ldots N \label{eq_KKT_slackness_mu}\\
-\breve{\xi}_k \breve{p}^B_k &= 0\;\text{for}\;k=1\ldots
N.\label{eq_KKT_slackness_xi}
\end{align}
where the partial derivatives in \eqref{eq_KKT_derivatives} are can
be expressed as
\vspace{-0.1cm}
\begin{eqnarray}
 \frac{\partial
\mathcal{L}_1}{\partial p^H_k} = - \tau_k \frac{\sqrt{\breve{p}^H_k}
+
\sqrt{\breve{p}^B_k}}{\sqrt{\breve{p}^H_k}\left(1+(\sqrt{\breve{p}^H_k}+\sqrt{\breve{p}^B_k})^2\right)}
+ \tau_k \sum_{n=k}^{N}  \breve{\lambda}_n - \breve{\mu}_k\label{eq.optimization_problem_proof_KKT_derivatives_xk}\nonumber\\
\frac{\partial \mathcal{L}_1}{\partial p^B_k}  = - \tau_k
\frac{\sqrt{\breve{p}^H_k} +
\sqrt{\breve{p}^B_k}}{\sqrt{\breve{p}^B_k}\left(1+(\sqrt{\breve{p}^H_k}+\sqrt{\breve{p}^B_k})^2\right)}
+ \tau_k \sum_{n=k}^{N}  \breve{\nu}_n -
\breve{\xi}_k\label{eq.optimization_problem_proof_KKT_derivatives_yk}\nonumber
\end{eqnarray}
From equation \eqref{eq_KKT_derivatives} and by introducing the
change of variables $\Breve{A}_k = \sum_{n=k}^{N} \breve{\lambda}_n
- \frac{\breve{\mu}_k}{\tau_k}$ and $\breve{B}_k = \sum_{n=k}^{N}
\breve{\nu}_n - \frac{\breve{\xi}_k}{\tau_k}$, the optimal transmit
powers in $k$-th epoch, $\breve{p}^H_k$  and $\breve{p}^B_k$, yield
\vspace{-0.1cm}
\begin{eqnarray}
 \breve{p}^H_k =
\frac{\breve{B}_k(\Breve{A}_k + \breve{B}_k - \Breve{A}_k
\breve{B}_k)}{\Breve{A}_k(\Breve{A}_k
+ \breve{B}_k)^2}\label{eq.optimization_problem_proof_KKT_solution_xk}\\
\breve{p}^B_k = \frac{\Breve{A}_k(\Breve{A}_k + \breve{B}_k -
\Breve{A}_k \breve{B}_k)}{\breve{B}_k(\Breve{A}_k +
\breve{B}_k)^2}.\label{eq.optimization_problem_proof_KKT_solution_yk}
\end{eqnarray}
Since, as stated in Lemma \ref{Lemma5} and equation
\eqref{eq_KKT_positivity_xy} above, $\breve{p}^H_k,\breve{p}^B_k
\neq 0$, the complementary slackness conditions
\eqref{eq_KKT_slackness_mu} and \eqref{eq_KKT_slackness_xi}, force
the corresponding Langrangian multipliers to vanish, i.e.
$\breve{\mu}_k, \breve{\xi}_k = 0$. When transmit power changes, we
have $\breve{p}^H_k < \breve{p}^H_{k+1}$. From
\eqref{eq.optimization_problem_proof_KKT_solution_xk} and the
\emph{Remark} below, this can only hold if $\breve{A}_k \neq
\breve{A}_{k+1}$ or, equivalently, if $\breve{\lambda}_{k} \neq 0$
(to recall, $\breve{\mu}_k=0$). From the complementary slackness
condition in \eqref{eq_KKT_slackness_lambda}, we have that
$\breve{\lambda}_{k} \neq 0 \Rightarrow \sum_{i=1}^k \tau_i
{\breve{p}^H_i} - E_k^{H}=0$. That is, the energy consumed by the
energy harvesting sensor up to $s_k$, equals the energy harvested by
such sensor up to that instant (see Fig.
\ref{fig.non_uniform_power_property_2}). This concludes the proof.

\emph{Remark}: From Lemma \ref{Lemma1}, we know that $\sum_{k=1}^{N}
\tau_k \breve{p}^B_k = E_0^2$. Since, in addition $\breve{p}^B_k
\neq 0$ this yields $\sum_{k=1}^{n} \tau_k \breve{p}^B_k - E_0^2<0$
for all $n=1 \ldots N-1$. From the complementary slackness condition
of \eqref{eq_KKT_slackness_nu}, we conclude that, necessarily,
$\breve{\nu}_k = 0$ for $k=1 \ldots N-1$. This, along with the fact
that $\breve{\mu}_k=0$ for all $k$, implies that $\breve{B}_k =
\breve{B}_N =\nu_N, \forall  k$, that is, all $\breve{B}_k$s are
identical. This property is a cornerstone of Algorithm
\ref{alg_battery} since it turns an $N$-dimensional exhaustive
search into a single-dimensional one.
\vspace{-0.3cm}
\section{Transmission policies with battery overflows are suboptimal}
\vspace{-0.1cm}
\begin{figure}[t]
   \centering
   \includegraphics[width=0.99\columnwidth]{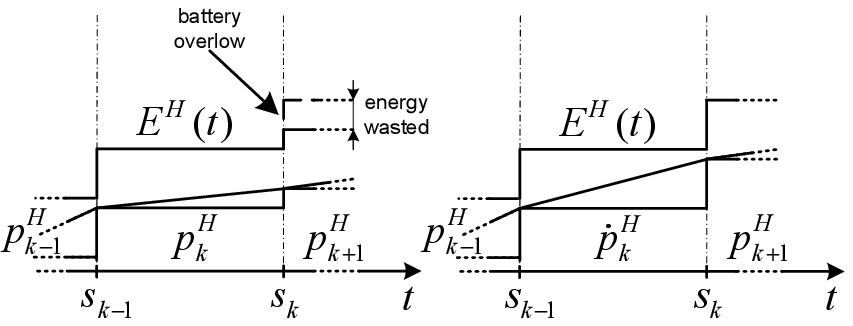}
      \vspace{-0.2cm}
   \caption{Transmission policies with battery overflows are strictly suboptimal.}
   \vspace{-0.3cm}
   \label{fig.Battery_overflow_suboptimal}
\end{figure}
Here we show that any transmission policy resulting into battery
overflows in the EH sensor is strictly suboptimal. We will prove
this by contradiction. Assume that a transmission policy with
battery overflow at $s_k$ only (Fig.
\ref{fig.Battery_overflow_suboptimal} left) is optimal. Let ${\Pi}^H
= \{p_1^H,\ldots,p_{k-1}^H, {p}_k^H,p_{k+1}^H,\dots, p_N^H\}$ and
${\Pi}^B = \{{p}_1^B,\ldots,{p}_{k-1}^B, {p}_k^B,{p}_{k+1}^B,\dots,
{p}_N^B\}$ denote the corresponding \emph{optimal} transmission
policies for the EH and BO sensors, respectively. We can think of an
alternative (and feasible) transmission policy
$\dot{\Pi}=\{\dot{\Pi}^H,\dot{\Pi}^B\}$ such that, on the one hand,
$\dot{\Pi}^H = \{p_1^H,\ldots,p_{k-1}^H,
\dot{p}_k^H,p_{k+1}^H,p_N^H\}$ and, on the other,
$\dot{\Pi}^B={\Pi}^B$. That is, the new policy only differs from the
optimal one in the power allocated to the EH sensor in the $k$-th
epoch. By properly adjusting $\dot{p}_k^H $, the battery overflow at
$s_k$ can be avoided (Fig. \ref{fig.Battery_overflow_suboptimal}
right). Since, clearly, $\dot{p}_k^H > {p}_k^H$, the throughput in
the $k$-th epoch is higher, this resulting into a higher total
throughput in $[0\ldots T]$. This contradicts the claim that the
original policy ${\Pi}=\{{\Pi}^H,{\Pi}^B\}$ is optimal and concludes
the proof.
\vspace{-0.3cm}
\section{Proof of Lemma \ref{Lemma7}}
\vspace{-0.1cm}
\label{sec.Proof_Lemma7} The Lagrangian $\mathcal{L}_2$ of the new
optimization problem with finite battery capacity constraints is
given by
\vspace{-0.1cm}
\begin{align}
\label{eq.optimization_problem_proof_lagrangian_lemma7}
\mathcal{L}_2 = &- \sum_{k=1}^{N} \tau_k \log \left(1+(\sqrt{p^H_k}
+
\sqrt{p^B_k})^2\right)\nonumber\\
\quad &+ \sum_{n=1}^{N} \lambda_n \left(\sum_{k=1}^{n} \tau_k p^H_k
- E_n^H\right) - \sum_{n=1}^{N} \pi_n  \left(\sum_{k=1}^n \tau_k
p_k^H -
E_n^{S}\right) \nonumber\\
\quad &+ \sum_{n=1}^{N} \nu_n \left(\sum_{k=1}^{n} \tau_k p^B_k -
E_n^B\right) - \sum_{k=1}^{N} \mu_k p^H_k - \sum_{k=1}^{N} \xi_k
p^B_k.
\end{align}
The new K.K.T. conditions thus read
\vspace{-0.1cm}
\begin{align}
\frac{\partial
\mathcal{L}_2}{\partial p^H_k}, \frac{\partial \mathcal{L}_2}{\partial p^B_k} &= 0\label{eq_KKT_derivatives_lemma7}\\
\sum_{k=1}^{n} \tau_k \breve{p}^H_k &\leq E_n^H \; \text{for}\;
n=1\ldots N \label{eq_KKT_causality_x_lemma7}\\
\sum_{k=1}^{n} \tau_k \breve{p}^H_k &\geq E_n^{S} \; \text{for}\;
n=1\ldots N \label{eq_KKT_battery_capacity_x_lemma7}\\
\sum_{k=1}^{n} \tau_k \breve{p}^B_k &\leq E_n^B \; \text{for}\;
n=1\ldots N \label{eq_KKT_causality_y_lemma7}\\
\breve{p}^H_k,\breve{p}^B_k &> 0\label{eq_KKT_positivity_xy_lemma7}\\
\breve{\lambda}_n, \breve{\pi}_n, \breve{\nu}_n, \breve{\mu}_k, \breve{\xi}_k &\geq 0 \label{eq_KKT_positivity_lagrangian_mult_lemma7}\\
\breve{\lambda}_n \left(\sum_{k=1}^{n} \tau_k \breve{p}^H_k - E_n^H\right) &= 0 \; \text{for}\; n=1\ldots N \label{eq_KKT_slackness_lambda_lemma7}\\
\breve{\pi}_n \left(\sum_{k=1}^{n} \tau_k \breve{p}^H_k - E_n^{S}\right) &= 0 \; \text{for}\; n=1\ldots N \label{eq_KKT_slackness_pi_lemma7}\\
\breve{\nu}_n \left(\sum_{k=1}^{n} \tau_k \breve{p}^B_k - E_n^B\right) &= 0 \; \text{for}\; n=1\ldots N \label{eq_KKT_slackness_nu_lemma7}\\
-\breve{\mu}_k \breve{p}^H_k &= 0 \; \text{for}\; k=1\ldots N \label{eq_KKT_slackness_mu_lemma7}\\
-\breve{\xi}_k \breve{p}^B_k &= 0\;\text{for}\;k=1\ldots N
\label{eq_KKT_slackness_xi_lemma7}
\end{align}
where equation \eqref{eq_KKT_slackness_pi_lemma7} accounts for the
additional constraint given by \eqref{eq.battery_constraint_1}, and
$\{\pi_n\}$ denote the corresponding set of Lagrange multipliers.
Since the additional constraint does not apply to the BO sensor, the
partial derivative $\frac{\partial \mathcal{L}_2}{\partial p^B_k}$
is identical to that in Appendix B, namely, $\frac{\partial
\mathcal{L}_2}{\partial p^B_k} = \frac{\partial
\mathcal{L}_1}{\partial p^B_k}$. On the contrary, $\frac{\partial
\mathcal{L}_2}{\partial p^H_k}$ differs and, more specifically, it
reads
\vspace{-0.1cm}
\begin{eqnarray}
 \frac{\partial
\mathcal{L}_2}{\partial p^H_k} = - \tau_k \frac{\sqrt{\breve{p}^H_k}
+
\sqrt{\breve{p}^B_k}}{\sqrt{\breve{p}^H_k}\left(1+(\sqrt{\breve{p}^H_k}+\sqrt{\breve{p}^B_k})^2\right)}
+ \tau_k (\sum_{n=k}^{N} \breve{\lambda}_n - \breve{\pi}_n)  -
\breve{\mu}_k\label{eq.optimization_problem_proof_KKT_derivatives_xk}
\end{eqnarray}
From \eqref{eq_KKT_derivatives_lemma7} and by introducing the change
of variables $\Breve{A}_k = \sum_{n=k}^{N} (\breve{\lambda}_n -
\breve{\pi}_n) - \frac{\breve{\mu}_k}{\tau_k}$ and $\breve{B}_k =
\sum_{n=k}^{N} \breve{\nu}_n - \frac{\breve{\xi}_k}{\tau_k}$, the
optimal transmit powers in $k$-th epoch, $\breve{p}^H_k$  and
$\breve{p}^B_k$, again yield
\vspace{-0.1cm}
\begin{eqnarray}
 \breve{p}^H_k =
\frac{\breve{B}_k(\Breve{A}_k + \breve{B}_k - \Breve{A}_k
\breve{B}_k)}{\Breve{A}_k(\Breve{A}_k
+ \breve{B}_k)^2}\label{eq.optimization_problem_proof_KKT_solution_xk_lemma7}\\
\breve{p}^B_k = \frac{\Breve{A}_k(\Breve{A}_k + \breve{B}_k -
\Breve{A}_k \breve{B}_k)}{\breve{B}_k(\Breve{A}_k +
\breve{B}_k)^2}.\label{eq.optimization_problem_proof_KKT_solution_yk_lemma7}
\end{eqnarray}
Equation \eqref{eq_KKT_positivity_xy_lemma7} and the complementary
slackness conditions \eqref{eq_KKT_slackness_mu_lemma7} and
\eqref{eq_KKT_slackness_xi_lemma7} again force the corresponding
Lagrangian multipliers to vanish, i.e. $\breve{\mu}_k, \breve{\xi}_k
= 0$. As in Appendix B, the transmit power changes ($\breve{p}^H_k
\neq \breve{p}^H_{k+1}$) iff $\breve{A}_k \neq \breve{A}_{k+1}$ or,
equivalently, if $\breve{\lambda}_k - \breve{\pi}_k \neq 0$. This is
only possible for the following combinations of values of the
Lagrangian multiplier: (i) $\breve{\lambda}_k \neq 0, \breve{\pi}_k
= 0$; (ii) $\breve{\lambda}_k = 0, \breve{\pi}_k \neq 0$; or (iii)
$\breve{\lambda}_k \neq 0, \breve{\pi}_k \neq 0, \breve{\lambda}_k
\neq  \breve{\pi}_k$. The conditions (i) and (ii) accounts for cases
in which the EC curve hits the cEH or cES curves at $s_k$
respectively; whereas (iii) accounts for the case in which the cEH
and cES curves coincide at time instant $s_k$ (i.e. when energy
harvested at $s_k$ equals battery capacity, namely, $E_k^1 =
E_{\text{max}}$).
\renewcommand{\thefigure}{\arabic{figure}}
\bibliographystyle{model1-num-names}
\bibliography{library_energy_harvesting}

\begin{thebibliography}{17}
\expandafter\ifx\csname natexlab\endcsname\relax\def\natexlab#1{#1}\fi
\providecommand{\bibinfo}[2]{#2}
\ifx\xfnm\relax \def\xfnm[#1]{\unskip,\space#1}\fi
\bibitem[{Yang and Ulukus(2012)}]{Yang_Ulukus_2011}
\bibinfo{author}{J.~Yang}, \bibinfo{author}{S.~Ulukus},
\newblock \bibinfo{title}{Optimal packet scheduling in an energy harvesting
  communication system},
\newblock \bibinfo{journal}{IEEE Transactions on Communications}
  \bibinfo{volume}{60} (\bibinfo{year}{2012}) \bibinfo{pages}{220 --230}.
\bibitem[{Tutuncuoglu and Yener(2012)}]{Tutuncuoglu_Yener_2012}
\bibinfo{author}{K.~Tutuncuoglu}, \bibinfo{author}{A.~Yener},
\newblock \bibinfo{title}{Optimum transmission policies for battery limited
  energy harvesting nodes},
\newblock \bibinfo{journal}{IEEE Transactions on Wireless Communications}
  \bibinfo{volume}{11} (\bibinfo{year}{2012}) \bibinfo{pages}{1180 --1189}.
\bibitem[{Ozel et~al.(2011)Ozel, Tutuncuoglu, Yang, Ulukus, and
  Yener}]{Ozel_Tutuncuoglu_2011_Journal}
\bibinfo{author}{O.~Ozel}, \bibinfo{author}{K.~Tutuncuoglu},
  \bibinfo{author}{J.~Yang}, \bibinfo{author}{S.~Ulukus},
  \bibinfo{author}{A.~Yener},
\newblock \bibinfo{title}{Transmission with energy harvesting nodes in fading
  wireless channels: Optimal policies},
\newblock \bibinfo{journal}{IEEE Journal on Selected Areas in Communications}
  \bibinfo{volume}{29} (\bibinfo{year}{2011}) \bibinfo{pages}{1732 --1743}.
\bibitem[{Yang and Ulukus(2012)}]{Yang2012_JCN}
\bibinfo{author}{J.~Yang}, \bibinfo{author}{S.~Ulukus},
\newblock \bibinfo{title}{Optimal packet scheduling in a multiple access
  channel with energy harvesting transmitters},
\newblock \bibinfo{journal}{Journal of Communications and Networks}
  \bibinfo{volume}{14} (\bibinfo{year}{2012}) \bibinfo{pages}{140 --150}.
\bibitem[{Tutuncuoglu and Yener(2012)}]{Tutuncuoglu_Yener_2011_JCN}
\bibinfo{author}{K.~Tutuncuoglu}, \bibinfo{author}{A.~Yener},
\newblock \bibinfo{title}{Sum-rate optimal power policies for energy harvesting
  transmitters in an interference channel},
\newblock \bibinfo{journal}{Journal of Communications and Networks}
  \bibinfo{volume}{14} (\bibinfo{year}{2012}) \bibinfo{pages}{151 --161}.
\bibitem[{Gunduz and Devillers(2011)}]{Gunduz_Devillers_2011}
\bibinfo{author}{D.~Gunduz}, \bibinfo{author}{B.~Devillers},
\newblock \bibinfo{title}{Two-hop communication with energy harvesting},
\newblock in: \bibinfo{booktitle}{4th IEEE International Workshop on
  Computational Advances in Multi-Sensor Adaptive Processing (CAMSAP), 2011},
  pp. \bibinfo{pages}{201 --204}.
\bibitem[{Yang et~al.(2012)Yang, Ozel, and Ulukus}]{Yang2012_ToWC}
\bibinfo{author}{J.~Yang}, \bibinfo{author}{O.~Ozel},
  \bibinfo{author}{S.~Ulukus},
\newblock \bibinfo{title}{Broadcasting with an energy harvesting rechargeable
  transmitter},
\newblock \bibinfo{journal}{Wireless Communications, IEEE Transactions on}
  \bibinfo{volume}{11} (\bibinfo{year}{2012}) \bibinfo{pages}{571 --583}.
\bibitem[{Ozel et~al.(2012)Ozel, Yang, and Ulukus}]{Ozel2012_ToWC}
\bibinfo{author}{O.~Ozel}, \bibinfo{author}{J.~Yang},
  \bibinfo{author}{S.~Ulukus},
\newblock \bibinfo{title}{Optimal broadcast scheduling for an energy harvesting
  rechargeable transmitter with a finite capacity battery},
\newblock \bibinfo{journal}{Wireless Communications, IEEE Transactions on}
  \bibinfo{volume}{11} (\bibinfo{year}{2012}) \bibinfo{pages}{2193 --2203}.
\bibitem[{Mudumbai et~al.(2010)Mudumbai, Hespanha, Madhow, and
  Barriac}]{Mudumbai2010}
\bibinfo{author}{R.~Mudumbai}, \bibinfo{author}{J.~Hespanha},
  \bibinfo{author}{U.~Madhow}, \bibinfo{author}{G.~Barriac},
\newblock \bibinfo{title}{Distributed transmit beamforming using feedback
  control},
\newblock \bibinfo{journal}{IEEE Transactions on Information Theory}
  \bibinfo{volume}{56} (\bibinfo{year}{2010}) \bibinfo{pages}{411 --426}.
\bibitem[{Pun et~al.(2008)Pun, Brown, and Vincent~Poor}]{Pun2009}
\bibinfo{author}{M.-O. Pun}, \bibinfo{author}{D.~Brown},
  \bibinfo{author}{H.~Vincent~Poor},
\newblock \bibinfo{title}{Opportunistic collaborative beamforming with one-bit
  feedback},
\newblock \bibinfo{journal}{SPAWC 2008.}  (\bibinfo{year}{2008})
  \bibinfo{pages}{246 --250}.
\bibitem[{Berbakov et~al.(2012)Berbakov, Matamoros, and
  Anton-Haro}]{Berbakov2012_ISWCS}
\bibinfo{author}{L.~Berbakov}, \bibinfo{author}{J.~Matamoros},
  \bibinfo{author}{C.~Anton-Haro},
\newblock \bibinfo{title}{Optimal transmission policy for distributed
  beamforming with energy harvesting and battery operated sensor nodes},
\newblock in: \bibinfo{booktitle}{9th International Symposium on Wireless
  Communication Systems (ISWCS) 2012}.
\bibitem[{Devillers and Gunduz(2012)}]{Devillers2012_JCN}
\bibinfo{author}{B.~Devillers}, \bibinfo{author}{D.~Gunduz},
\newblock \bibinfo{title}{A general framework for the optimization of energy
  harvesting communication systems with battery imperfections},
\newblock \bibinfo{journal}{Communications and Networks, Journal of}
  \bibinfo{volume}{14} (\bibinfo{year}{2012}) \bibinfo{pages}{130 --139}.
\bibitem[{Burden and Faires(2011)}]{Burden_Numerical_Analysis}
\bibinfo{author}{R.~L. Burden}, \bibinfo{author}{J.~D. Faires},
  \bibinfo{title}{Numerical Analysis}, \bibinfo{publisher}{Brooks/Cole},
  \bibinfo{year}{2011}.
\bibitem[{Saggini et~al.(2010)Saggini, Ongaro, Galperti, and
  Mattavelli}]{Saggini_supercap_storage}
\bibinfo{author}{S.~Saggini}, \bibinfo{author}{F.~Ongaro},
  \bibinfo{author}{C.~Galperti}, \bibinfo{author}{P.~Mattavelli},
\newblock \bibinfo{title}{Supercapacitor-based hybrid storage systems for
  energy harvesting in wireless sensor networks},
\newblock in: \bibinfo{booktitle}{Applied Power Electronics Conference and
  Exposition (APEC), 2010 Twenty-Fifth Annual IEEE}, pp. \bibinfo{pages}{2281
  --2287}.
\bibitem[{Li et~al.(2001)Li, Murphy, Winnick, and
  Kohl}]{Li_pulse_charging_lion_batteries}
\bibinfo{author}{J.~Li}, \bibinfo{author}{E.~Murphy},
  \bibinfo{author}{J.~Winnick}, \bibinfo{author}{P.~A. Kohl},
\newblock \bibinfo{title}{The effects of pulse charging on cycling
  characteristics of commercial lithium-ion batteries},
\newblock \bibinfo{journal}{Journal of Power Sources} \bibinfo{volume}{102}
  (\bibinfo{year}{2001}) \bibinfo{pages}{302 -- 309}.
\bibitem[{NSR(2005)}]{NSRDB}
\bibinfo{title}{National solar radiation data base}, \bibinfo{year}{1991-2005}.
  \bibinfo{note}{\url{http://rredc.nrel.gov/solar/old_data/nsrdb/1991-2005/tmy%
3/}}.
\bibitem[{Polyanin and Manzhirov(2007)}]{Polyanin_Manizhirov_Math}
\bibinfo{author}{A.~D. Polyanin}, \bibinfo{author}{A.~V. Manzhirov},
  \bibinfo{title}{Handbook of Mathematics for Engineers and Scientists},
  \bibinfo{publisher}{Chapman \& Hall/CRC}, \bibinfo{year}{2007}.

\end{thebibliography}






\end{document}